\documentclass[twocolumn,prl,aps,superscriptaddress]{revtex4}

\usepackage{xcolor}
\usepackage{epsfig}
\usepackage{graphicx}
\usepackage{dcolumn}
\usepackage{bm} 
\usepackage{epstopdf}
\usepackage{amsmath}
\usepackage{amsfonts}
\usepackage[colorlinks,linkcolor=blue,anchorcolor=blue,citecolor=blue]{hyperref}
\newcommand{\beq}{\begin{equation}}
\newcommand{\eeq}{\end{equation}}
\newcommand{\beqa}{\begin{eqnarray}}
\newcommand{\eeqa}{\end{eqnarray}}

\newcommand{\om}{\omega}

\def\jpb#1{{ J.\ Phys.\ B} {\bf#1}}

\def\prc#1{{ Phys.\ Rev. C\/} {\bf#1}}

\def\prl#1{{ Phys.\ Rev.\ Lett.} {\bf#1}}

\begin{document}

\title{Recollision induced nuclear excitation of $^{229}$Th}

\author{Xu Wang}
\email{xwang@gscaep.ac.cn}
\affiliation{Graduate School, China Academy of Engineering Physics, Beijing 100193, China}

\date{\today}

\begin{abstract}

Previously we proposed a new approach of exciting the $^{229}$Th nucleus using laser-driven electron recollision [W. Wang et al., Phys. Rev. Lett. {\bf 127}, 052501 (2021)]. The current article is aimed to elaborate the method by explaining further theoretical details and presenting extended new results. The method has also been improved by adopting the electronic excitation cross sections calculated recently by Tkalya [E. V. Tkalya, Phys. Rev. Lett. {\bf 124}, 242501 (2020)]. The new cross sections are obtained from Dirac distorted-wave calculations instead of from Dirac plane-wave calculations as we used previously. The distorted-wave cross sections are shown to be 5 to 6 orders of magnitude higher than the plane-wave results. With the excitation cross sections updated, the probability of isomeric excitation of $^{229}$Th from electron recollision is calculated to be on the order of $10^{-12}$ per nucleus per (femtosecond) laser pulse. Dependency of the excitation probability on various laser parameters is calculated and discussed, including the laser intensity, the laser wavelength, and the laser pulse duration.

\end{abstract}

\maketitle

\section{1. Introduction}

The $^{229}$Th nucleus has a unique low-lying isomeric state of energy (currently known as) around 8.3 eV above the nuclear ground state \cite{Kroger-76, Reich-90, Helmer-94, Beck-07, Seiferle-19}. This isomeric state is the lowest nuclear excited state so far known, and its existence has fascinated the scientific community for its potential applications in nuclear optical clocks \cite{Peik-03, Peik-09, Rellergert-10, Campbell-12}, in nuclear lasers \cite{Tkalya-11}, in checking variations of fundamental constants \cite{Flambaum-06, Berengut-09, Fadeev-20}, etc.

The isomeric state can be obtained from $\alpha$ decay of $^{233}$U ($^{233}$U $\rightarrow$ $^{229}$Th + $\alpha$, half-life about 1.6 $\times 10^5$ years, with 2\% of the resultant $^{229}$Th nuclei in the isomeric state), although the efficiency is rather low. One can estimate that every $3.6\times 10^{14}$ $^{233}$U nuclei generate a single $^{229}$Th nucleus in the isomeric state per second. Besides, the $^{229}$Th nucleus is left with a recoil energy of 84 keV into random directions and various ionic states. To realize the above-mentioned applications, {\it controllable and efficient} excitation of the $^{229}$Th nucleus is needed, yet it remains a major problem to be solved.

The following approaches have been attempted experimentally or proposed theoretically to excite the $^{229}$Th nucleus: 

(i) Direct light excitation. Vacuum-ultraviolet (VUV) light sources around 8 eV photon energies can be generated from synchrotron radiations and from frequency combs (high harmonic generation). Several experimental attempts have been made to radiate $^{229}$Th nuclei with VUV lights and detect the subsequent fluorescence \cite{Jeet-15, Yamaguchi-15, Peik-15, Stellmer-18}. However, no positive results have been reported observing the fluorescence signals with the desired lifetime characteristics. 

(ii) Indirect light excitation. Masuda et al. use 29 keV synchrotron radiations to excite the $^{229}$Th nuclei to the second excited state which then decays predominantly to the isomeric state \cite{Masuda-19}. The probability of excitation to the isomeric state for a single $^{229}$Th nucleus is estimated to be on the order of $10^{-11}$ per second. Up to now this is the only experimentally realized excitation of the $^{229}$Th nucleus. This approach, however, requires narrowband high-photon-energy synchrotron light sources which are not easily accessible.

(iii) Electronic bridge (EB) excitation schemes. The idea is to couple the nuclear and the electronic degrees of freedom and to use the energy released from an electronic transition to excite the nucleus. An additional laser, which is presumably easily accessible, is used to compensate the energy mismatch between the electronic transition and the nuclear transition. Several ionic or doped-crystal schemes have been proposed \cite{Porsev-10, Borisyuk-19, Bilous-20, Nickerson-20}. The EB approach requires accurate knowledge of both the isomeric energy and the electronic structures of the $^{229}$Th ions. Experimental realizations of the EB schemes have not been reported.   

Previously we proposed a new excitation approach using laser-driven electron recollision \cite{Wang-21}. The approach was termed recollision induced nuclear excitation (RINE). Recollision \cite{Kulander-93, Schafer-93, Corkum-93} is the core process of strong-field atomic physics, and it is the underlying mechanism of various strong-field phenomena including high harmonic generation \cite{McPherson-87, Ferray-88, Seres-05}, attosecond pulse generation \cite{Krausz-09, Zhao-12, Li-17, Gaumnitz-17}, nonsequential double ionization \cite{Walker-94, Palaniyappan-05, Becker-12}, laser-induced electron diffraction \cite{Morishita-08, Blaga-12, Wolter-16}, etc. Recolliding electrons usually have energy up to several tens of electronvolts, or possibly up to several hundreds of electronvolts with substantially reduced fluxes, so they usually do not affect the nucleus. However, for $^{229}$Th the recolliding electrons do have enough energy to excite the nucleus to the isomeric state. The RINE approach is therefore the result of a combination of strong-field atomic physics and $^{229}$Th nuclear physics \cite{Wang-21jpb}.    

The goal of the current article is to elaborate the RINE approach by explaining further details of the method itself and presenting extended new results. The method will also be improved by updating the electronic excitation cross sections from the Dirac plane-wave results \cite{Alder-56} to the Dirac distorted-wave results recently calculated by Tkalya \cite{Tkalya-20}. The distorted-wave cross sections are shown to be {\it 5 to 6 orders of magnitude higher} than the plane-wave results. With the updated excitation cross sections, the RINE approach is shown to be very efficient: the probability of isomeric excitation for a $^{229}$Th nucleus is calculated to be on the order of $10^{-12}$ per laser pulse (with duration $\sim 10$ fs).

This article is organized as follows. In Section 2 the RINE method is explained in detail by examining each of the involved theoretical elements. In Section 3 numerical results are presented, including the dependency of the nuclear excitation probability on the laser intensity, the laser pulse duration, and the laser wavelength. Further discussions and remarks are given in Section 4. A conclusion is given in Section 5.

\section{2. The RINE method}

\subsection{2.1 Overview}

\begin{figure} [t!]
 \centering
 \includegraphics[width=7cm, trim=0 0 0 0]{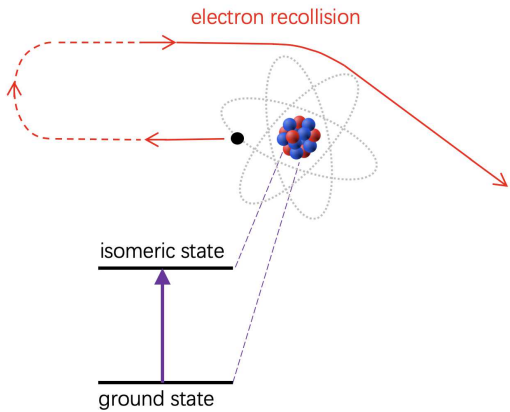} 
 \caption{Illustration of the RINE approach. An outer electron of the $^{229}$Th atom (or ion) is emitted into the continuum via tunneling ionization. It is later driven back by the oscillating laser electric field, recollides with its parent ion core, and excites the nucleus from the ground state to the isomeric state.}\label{f.illustration}
\end{figure}

The idea of the RINE approach is illustrated in Fig. \ref{f.illustration}. In a strong laser field (typically of peak intensity on the order of $10^{13}$ to $10^{15}$ W/cm$^2$), a $^{229}$Th atom (or ion) is ionized via quantum tunneling. The emitted electron, albeit in the continuum, has a probability of being driven back to collide with its parent ion core when the oscillating laser electric field reverses its direction. This is called a recollision process. The $^{229}$Th nucleus has a probability of being excited to the isomeric state by the recolliding electron. If the laser field is strong enough, several electrons may be pulled out and driven to recollide, at different time intervals during the laser pulse though.

Let us now consider a single electron, which could be the first- (second-, third-, fourth-, ...) emitted electron. Of course, different electrons have different ionization energies and see different potentials from the remaining ion core. At each time $t_i$ during the laser pulse, the electron can be emitted via tunneling ionization with a rate $w(t_i)$. The emitted electron may be driven back and recollide with its parent ion core at a later time $t_r$. The effective flux density of the recolliding electron is given by
\beq
j(t_r) = \frac{w(t_i) dt_i P(t_i,R_c)}{\pi R_c^2 d t_r}. \label{e.jtr}
\eeq
This formula is understood as follows: $w(t_i) dt_i$ is the probability of tunneling ionization at time $t_i$ within a small time interval $dt_i$. A fraction $0 \le P(t_i,R_c) <1$ of this probability will experience recollision {\it and} recollide within a critical radius $R_c$ from the nucleus. Only this fraction of the recollision events contribute to nuclear excitation, as will be explained later in Section 2.3. The contributing probability is then divided by the area $\pi R_c^2$ and the recolliding time interval $dt_r$ to give the effective flux density at time $t_r$. Note that $R_c$ has a weak dependency on the recollision energy $E_r$, as will be shown in Section 2.3, so it is a function of the recollision time, i.e. $R_c = R_c(t_r)$.

The nuclear excitation rate at time $t_r$ is given by
\beq
\Gamma_\text{exc} (t_r) = \sigma(E_r) j(t_r) \beta(t_r), \label{e.Gamma}
\eeq   
where $E_r$ is the energy of the electron at $t_r$, $\sigma(E_r)$ is the corresponding nuclear excitation cross section, and $j(t_r)$ is the effective flux density of the recolliding electron. The factor 
\beq
\beta(t_r) = \frac{R_c^2(t_r)}{b_c^2(t_r)} \label{e.beta}
\eeq 
transforms the recollision-plane (viz. the $x=0$ plane) flux density to the corresponding asymptotic flux density. The reason to perform such a transformation is that we are using the cross section $\sigma(E_r)$ obtained for an electron wave coming from infinity. A recolliding electron flux with cross area $\pi R_c^2$ at the recollision plane {\it comes as if} from infinity with cross area $\pi b_c^2$. Here $b_c$ is the impact parameter corresponding to $R_c$. However, as will be shown later in Section 2.3, $\beta(t_r)$ is slightly smaller than 1, so omitting this $\beta$ factor will not affect the excitation rate substantially. 

The probability of nuclear excitation is obtained by a time integral of the excitation rate
\beqa
P_\text{exc}(t) &=& \int_{-\infty}^{t} \Gamma_\text{exc}  (t_r) dt_r \nonumber \\
&=& \int_{-\infty}^{t} \sigma(E_r) \frac{w(t_i) P(t_i,R_c)}{\pi b_c^2} dt_i, \label{e.Pexc}
\eeqa
where Eqs. (\ref{e.jtr} - \ref{e.beta}) have been substituted into the first line to get the second line. 
Elements involved in the above formula, such as the electronic excitation cross section $\sigma(E_r)$, the ionization rate $w(t_i)$, the critical recollision radius $R_c$ and the corresponding impact parameter $b_c$, and the probability $P(t_i, R_c)$, will be explained in the following subsections.

\subsection{2.2 The electronic excitation cross section}

\begin{figure} [t!]
 \centering
 \includegraphics[width=7cm, trim=0 0 0 0]{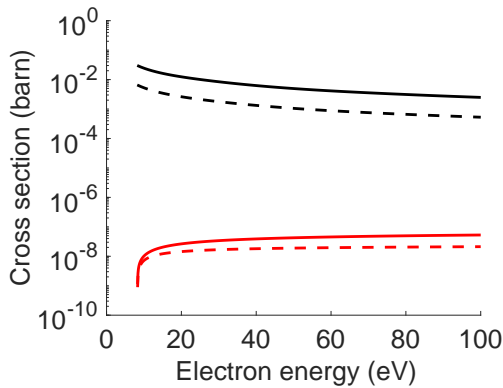} 
 \caption{Total ($E2 + M1$) electronic excitation cross sections of $^{229}$Th from the ground state to the isomeric state. The black (upper two) curves are from distorted-wave calculations and the red (lower two) curves are from plane-wave calculations. For each calculation, the solid curve uses set 1 (Eqs. \ref{e.BM1set1} - \ref{e.BE2set1}) of the reduced transition probabilities and the dashed curve uses set 2 (Eqs. \ref{e.BM1set2} - \ref{e.BE2set2}).} \label{f.crosec}
\end{figure}

In nuclear physics, Coulomb excitation is a very useful method for the study of nuclear structures, especially collective levels from rotational or vibrational degrees of freedom \cite{Alder-56, Winther-79, Cline-86, Motobayashi-95, Gorgen-16, Feng-22}. A beam of projectile particles bombard on and excite the target nuclei through the mutual Coulomb interaction. Commonly used projectiles in Coulomb-excitation experiments are protons and $\alpha$ particles, whereas in the current article the case of electrons is considered. The target is the $^{229}$Th nucleus. 

From the theoretical side, relatively simple analytical formulas are available for the excitation cross sections if the electrons are treated as (Dirac) plane waves, i.e. with the plane-wave Born approximation. For the case of $^{229}$Th, the dominant nuclear transition between the ground state and the isomeric state is electric quadrupole ($E2$) and magnetic dipole ($M1$). (The ground state has spin and parity $5/2^+$, and the isomeric state has spin and parity $3/2^+$. Electric dipole transition is forbidden.) From Ref. \cite{Alder-56}, the differential excitation cross sections are given by 
\beqa
\frac{d\sigma_{E2}}{d\Omega} &=& B(E2; g \rightarrow e) \frac{2\pi}{75c^2}  \frac{K^4}{k_i^2} \left( V_T + \frac{2}{3} V_L \right) ,  \label{e.DsigmaE2pl} \\
\frac{d\sigma_{M1}}{d\Omega} &=& B(M1; g \rightarrow e) \frac{8\pi}{9c^2}  \frac{K^2}{k_i^2} V_T .  \label{e.DsigmaM1pl}
\eeqa
The total excitation cross sections can be obtained after integrating over the solid angle. In the above formulas $c$ is the speed of light, $\bm{k}_i$ ($\bm{k}_f$) is the initial (final) wave vector of the electron, and $\bm{K} = \bm{k}_i - \bm{k}_f$ is the momentum transfer. $V_T$ and $V_L$ are shorthand notations of
\beqa
V_T &=& k_i k_f \frac{(k_i^2 + k_f^2 -\kappa^2)K^2-2(\bm{k}_i \cdot \bm{K})(\bm{k}_f \cdot \bm{K})}{K^2(K^2-\kappa^2)^2}, \\
V_L &=&  k_i k_f \frac{2 k_i^2 + 2 k_f^2 + 4c^2 - \kappa^2 - K^2}{K^4},
\eeqa
where $\kappa = \Delta E/c$ with $\Delta E$ the energy transfer (i.e. the energy difference between the isomeric state and the ground state, taken to be 8.3 eV in the current article). 

$B(E2; g \rightarrow e)$ and $B(M1; g \rightarrow e)$ are the reduced transition probabilities, and the notation $g \rightarrow e$ means from the nuclear ground state to the isomeric excited state. The following relation holds if the transition direction is reversed
\beqa
 \frac{B(E2/M1; g \rightarrow e)}{B(E2/M1; e \rightarrow g)} = \frac{2 I_e + 1}{2 I_g + 1},
\eeqa
where $I_g = 5/2$ and $I_e = 3/2$ are the nuclear spin for the ground state and for the isomeric state. The values of the reduced transition probabilities are determined either from analyses of $\gamma$-ray spectra of excited $^{229}$Th nuclei exploiting Alaga rules \cite{Bemis-88, Dykhne-98, Gulda-02, Barci-03, Ruchowska-06, Tkalya-15} or from nuclear model calculations \cite{Barci-03, Ruchowska-06, Minkov-17, Minkov-21}. There are some degrees of uncertainties with these values at the current stage. For example, Ref. \cite{Dykhne-98} suggests $B(M1; e \rightarrow g)$ = 0.048 W.u. (Weisskopf units), Ref. \cite{Ruchowska-06} suggests $B(M1; e \rightarrow g)$ = 0.014 W.u. and $B(E2; e \rightarrow g)$ = 67 W.u., Ref. \cite{Minkov-17} suggests $B(M1; e \rightarrow g)$ = 0.0076 W.u. and $B(E2; e \rightarrow g)$ = 27 W.u., and Ref. \cite{Minkov-21} suggests $B(M1; e \rightarrow g)$ to be between 0.005 and 0.008 W.u. and $B(E2; e \rightarrow g)$ to be between 30 and 50 W.u. First-principle many-body nuclear calculations are not expected to be available in the near future.

In Ref. \cite{Tkalya-20} Tkalya compares the excitation cross sections with the following two sets of reduced transition probabilities (with the $g \rightarrow e$ values converted to $e \rightarrow g$ values): 
\beqa
\text{set 1: }  B(M1; e \rightarrow g) &=& 0.048 \text{ W.u.}  \label{e.BM1set1}\\
                     B(E2; e \rightarrow g) &=& 17.6 \text{ W.u.}    \label{e.BE2set1}\\
\text{set 2: }  B(M1; e \rightarrow g) &=& 0.0076 \text{ W.u.}  \label{e.BM1set2} \\
                     B(E2; e \rightarrow g) &=& 27 \text{ W.u.}         \label{e.BE2set2}
\eeqa
With these two sets of the reduced transition probabilities and Eqs. (\ref{e.DsigmaE2pl}-\ref{e.DsigmaM1pl}), we can get the total ($E2 + M1$) electronic excitation cross sections, as shown in the red (lower two) curves of Fig. \ref{f.crosec}. The solid curve is for set 1 and the dashed curve is for set 2. The two curves are within a factor of 2 or 3. We have checked that other suggested values of the reduced transition probabilities mentioned above lead to cross sections roughly within the range of set 1 and set 2. One also sees that the plane-wave formulas result in excitation cross sections on the order of $10^{-8}$ barn, or $10^{-32}$ cm$^2$.

Recently Tkalya calculated the electronic excitation cross sections using Dirac distorted waves \cite{Tkalya-20} instead of plane waves. That is, he used the distorted-wave Born approximation. The results show that the excitation cross sections are 5 to 6 orders of magnitude higher than the plane-wave values, as shown in Fig. \ref{f.crosec} by the black (upper two) curves (solid for set 1 and dashed for set 2 of the reduced transition probabilities). The cross sections are on the order of $10^{-3}$ to $10^{-2}$ barn. Unlike the plane-wave cross sections which ignore the ion-core potential, the distorted-wave cross sections depend on the ion-core potential, although the dependency is rather weak. The black curves shown in Fig. \ref{f.crosec} are for the $^{229}$Th$^+$ ion, but the cross sections for the $^{229}$Th$^{2+}$, $^{229}$Th$^{3+}$, and $^{229}$Th$^{4+}$ ions are almost visually indistinguishable from the $^{229}$Th$^+$ case. The cross sections for the neutral $^{229}$Th atom are a little different though \cite{Tkalya-20, Zhang-22}, however for the RINE method, only cross sections for the first few ions are of concern.

The surprising, but not totally unexpected, difference between the distorted-wave results and the plane-wave results stems mainly from the fact that the electron energies considered here are very low (mostly below 100 eV), so plane waves turn out to be bad approximations to the actual wave functions of the electron. Except for a couple of minor and insignificant errors (including an overall factor of 2 larger possibly from summing over angular indices, and a confusion of the nuclear transition direction of set 2 of the reduced transition probabilities), our independent calculations \cite{Zhang-22} confirm the results of Tkalya \cite{Tkalya-20}. (The cross sections shown in Fig. \ref{f.crosec} have had the errors corrected.) Detailed formulas of the Dirac distorted-wave calculations can be found in \cite{Tkalya-20} and will not be repeated here.

In our previous article \cite{Wang-21}, the plane-wave cross sections were used. In the current article, the distorted-wave cross sections will be adopted. As expected, the nuclear excitation probabilities reported in the current article are roughly 5 to 6 orders of magnitude higher than the results reported in \cite{Wang-21}.

\subsection{2.3 The adiabaticity of collision trajectories and the effective collision area}

\begin{figure} [t!]
 \centering
 \includegraphics[width=4.2cm, trim=0 0 0 0]{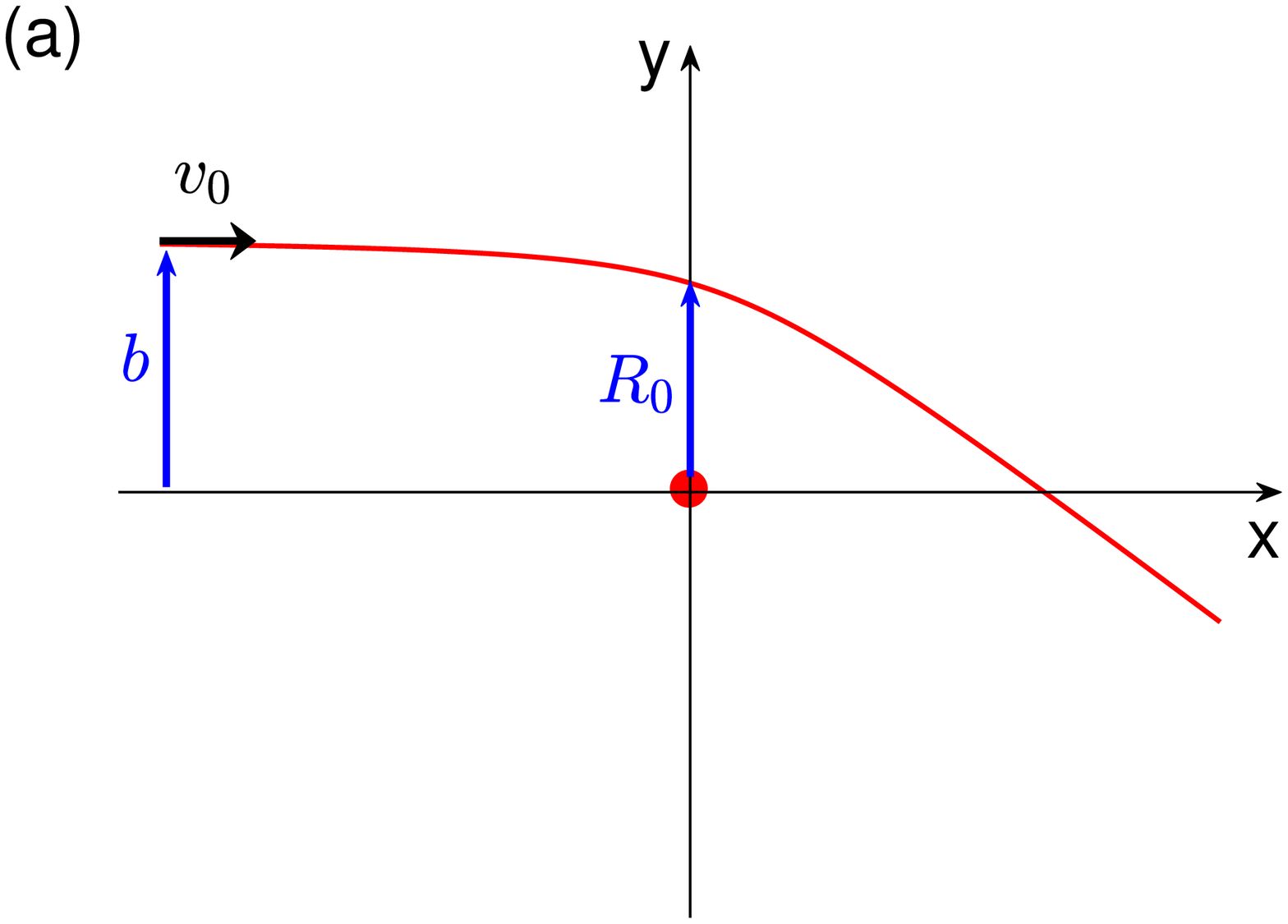} 
 \includegraphics[width=4.2cm, trim=0 0 0 0]{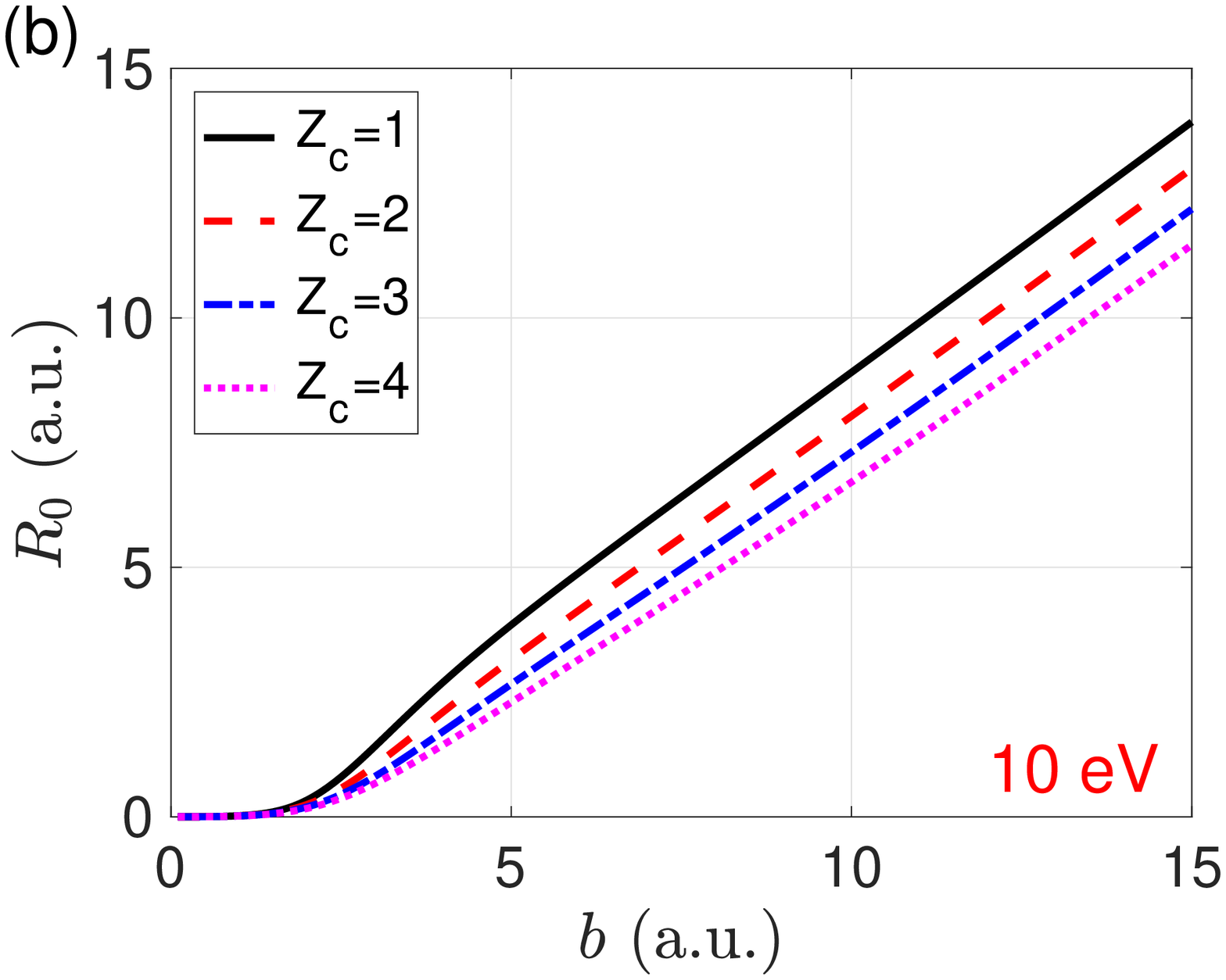} \\
 \vspace{0.1cm}
 \includegraphics[width=4.2cm, trim=0 0 0 0]{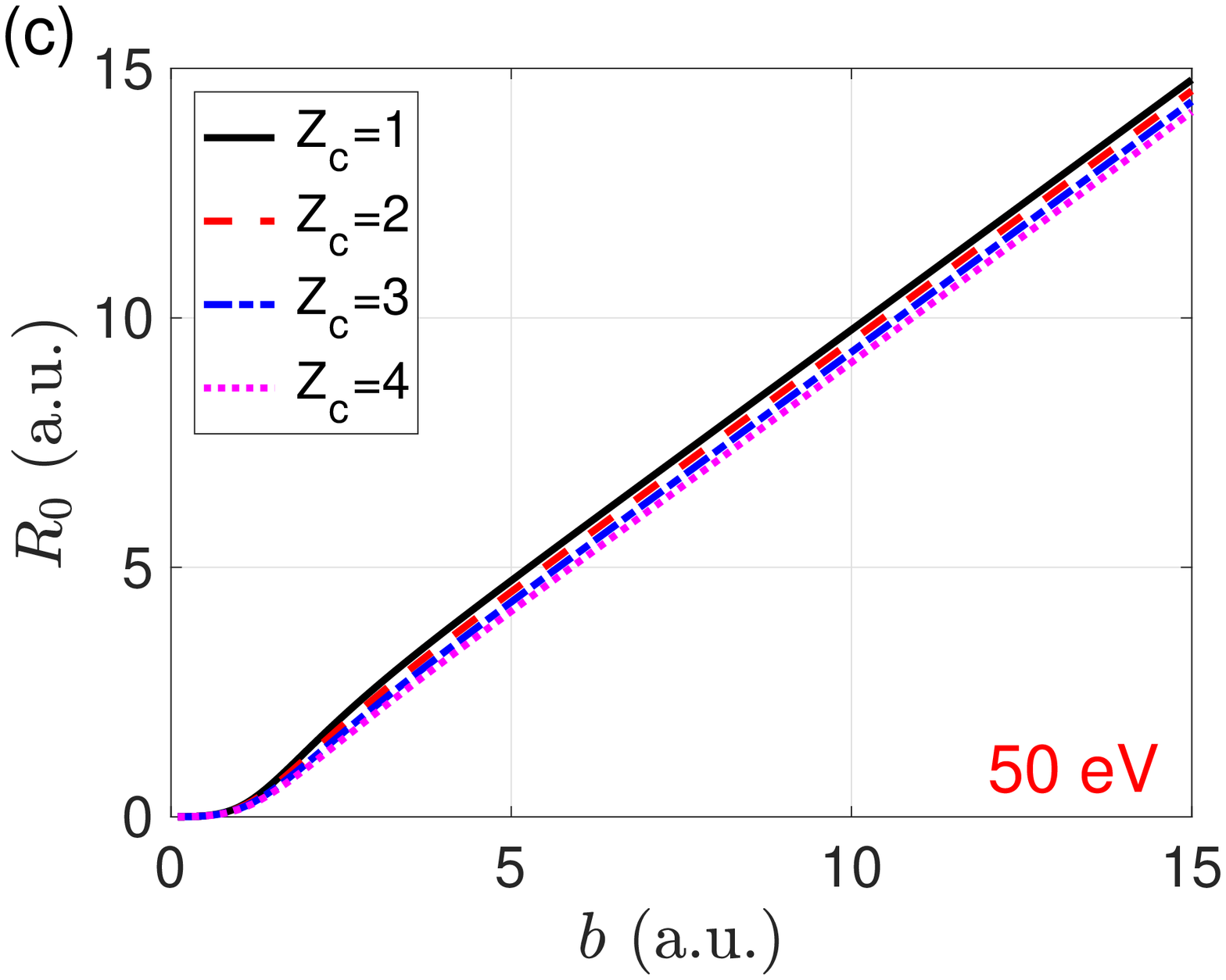} 
 \includegraphics[width=4.2cm, trim=0 0 0 0]{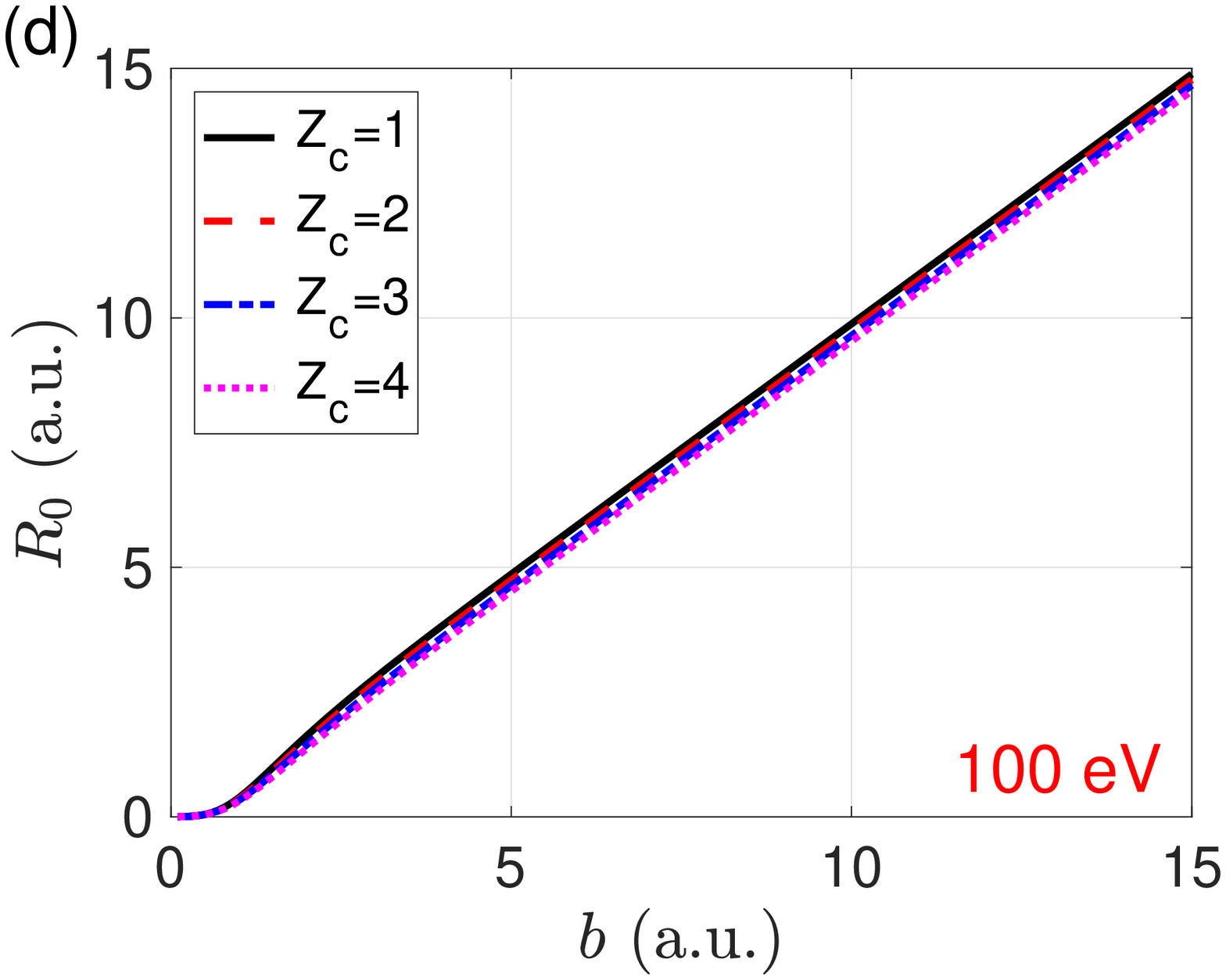} 
 \caption{(a) Illustration of an electron collision trajectory. The electron has an asymptotic velocity $v_0$ and an impact parameter $b$. Its distance to the nucleus is denoted $R_0$ as the electron passes the $x=0$ plane. (b-d) The relationship between $R_0$ and $b$ for electron energies 10, 50, and 100 eV, as labeled on each figure. For each energy, four different ion-core charges are used. The ion-core potential is described by the GSZ potential (Eq. \ref{e.GSZ}).}\label{f.bR0}
\end{figure}

\begin{figure} [b!]
 \centering
 \includegraphics[width=4.2cm, trim=0 0 0 0]{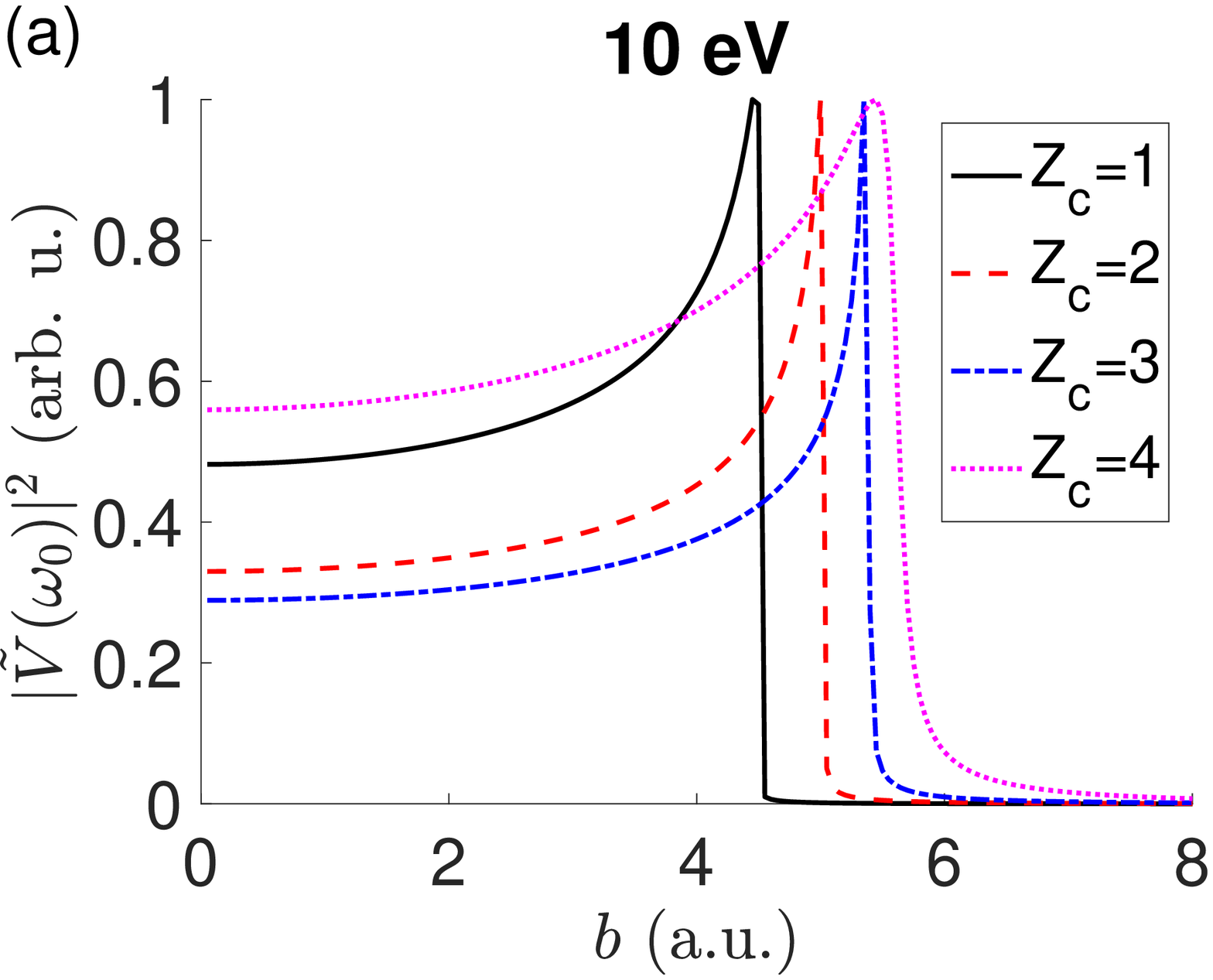} 
 \includegraphics[width=4.2cm, trim=0 0 0 0]{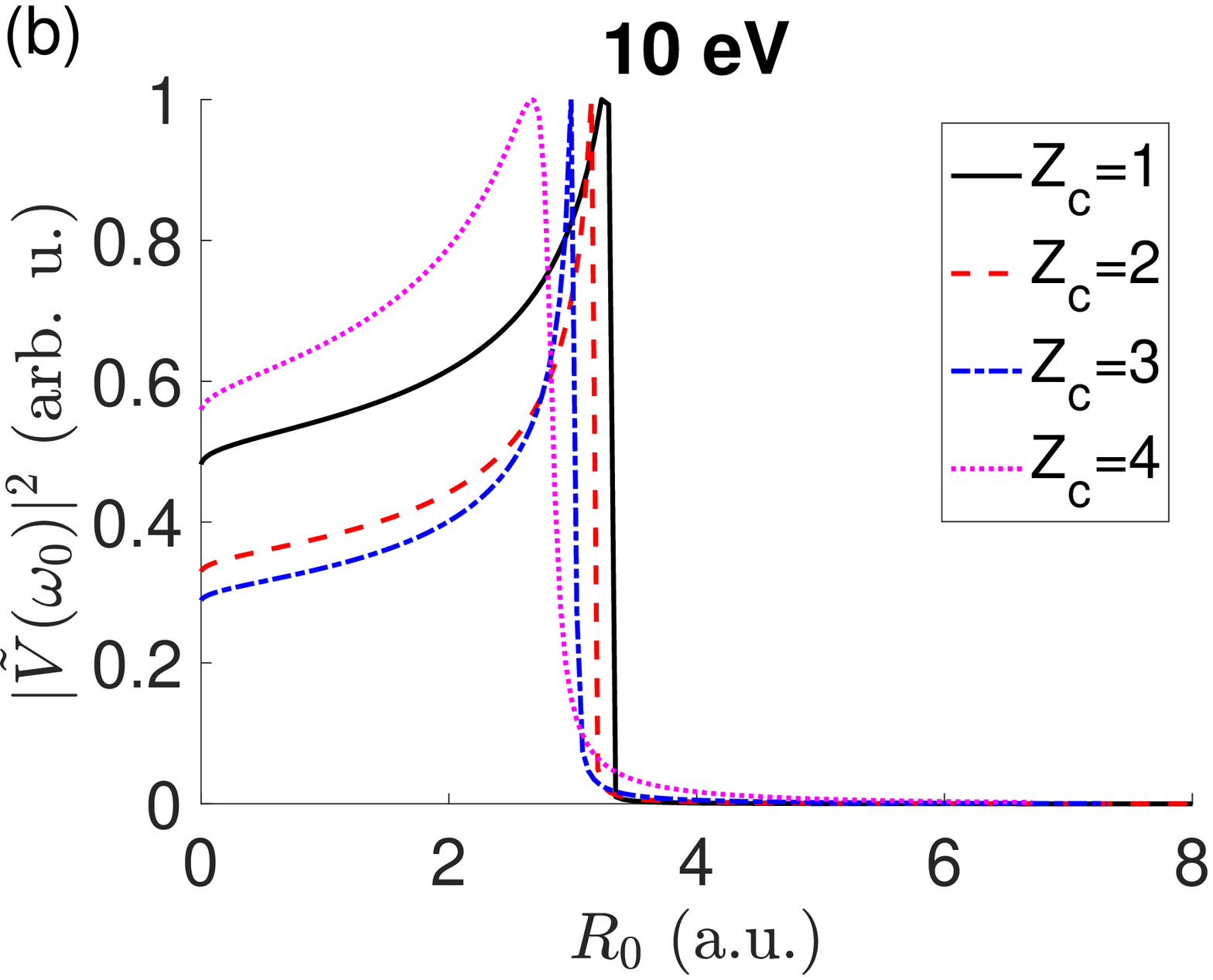} \\
 \vspace{0.2cm}
 \includegraphics[width=4.2cm, trim=0 0 0 0]{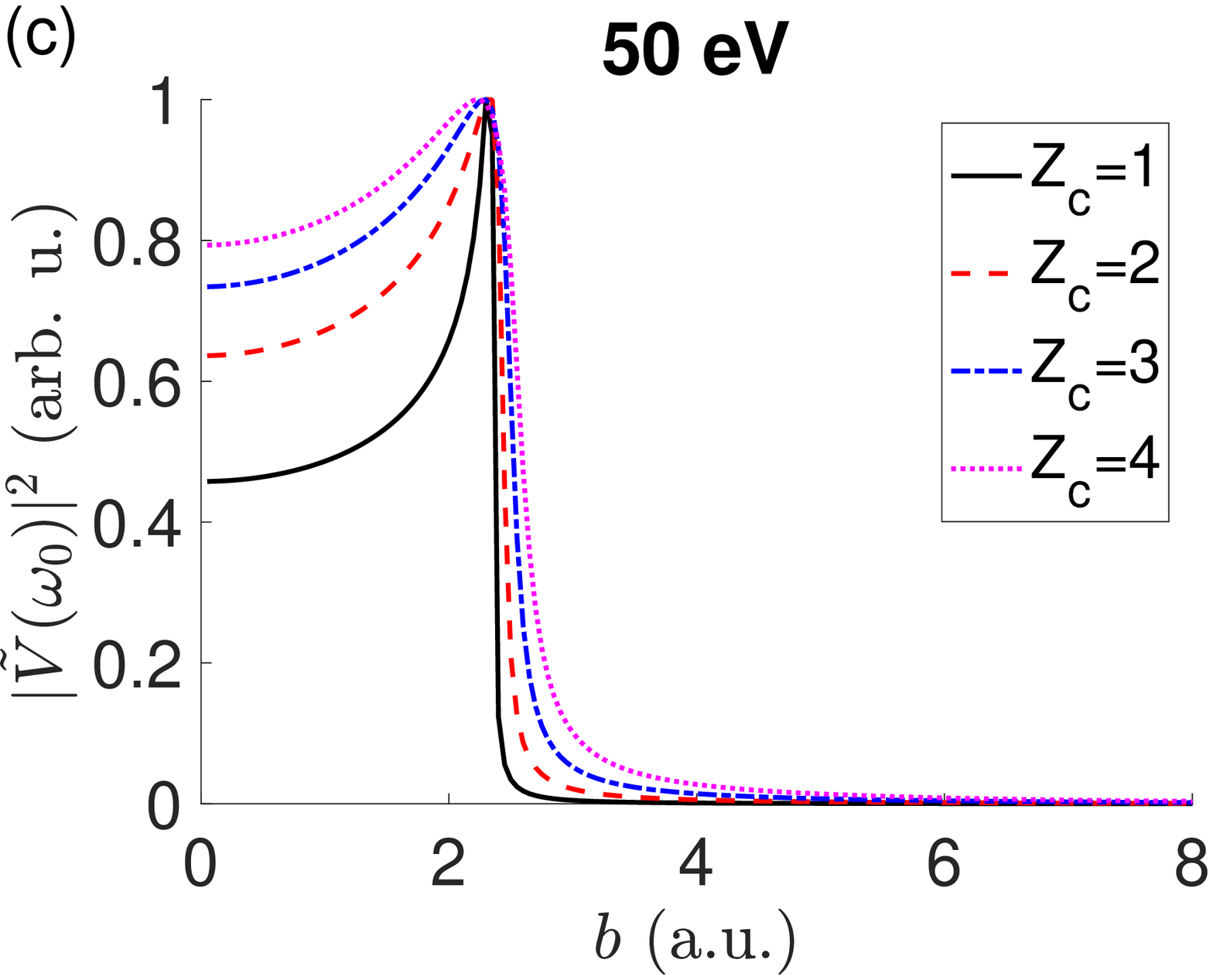}
 \includegraphics[width=4.2cm, trim=0 0 0 0]{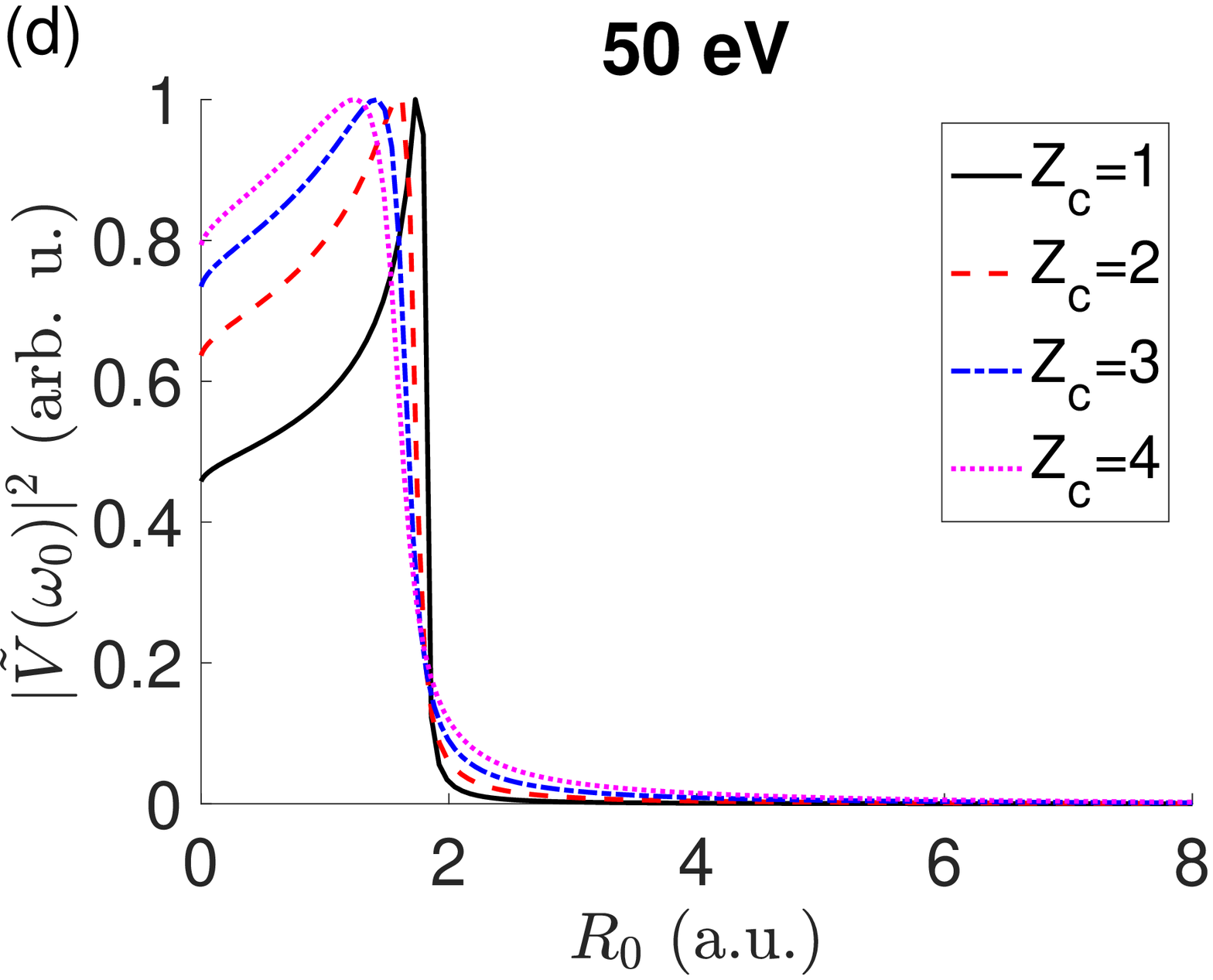}\\
 \vspace{0.2cm}
 \includegraphics[width=4.2cm, trim=0 0 0 0]{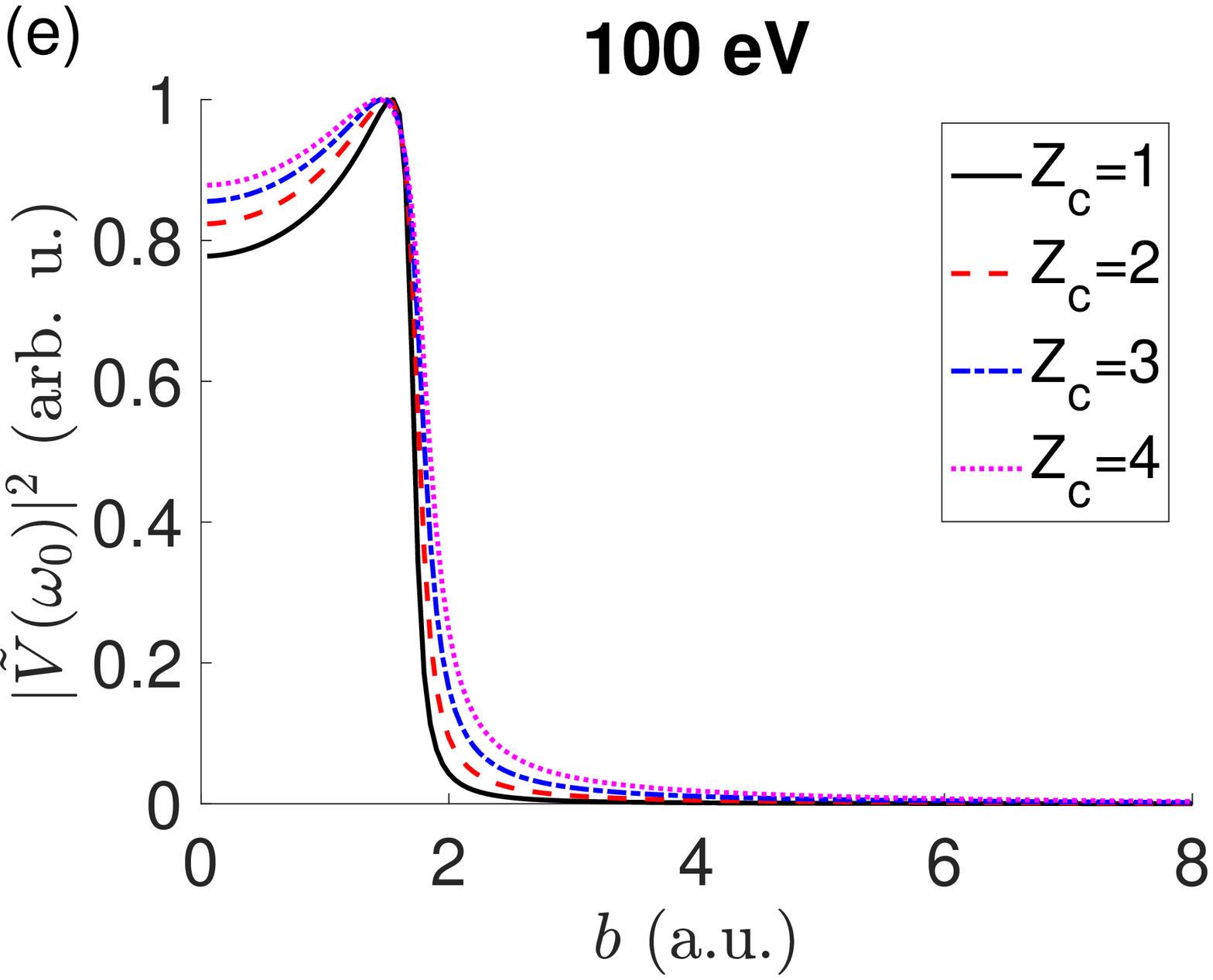}
 \includegraphics[width=4.2cm, trim=0 0 0 0]{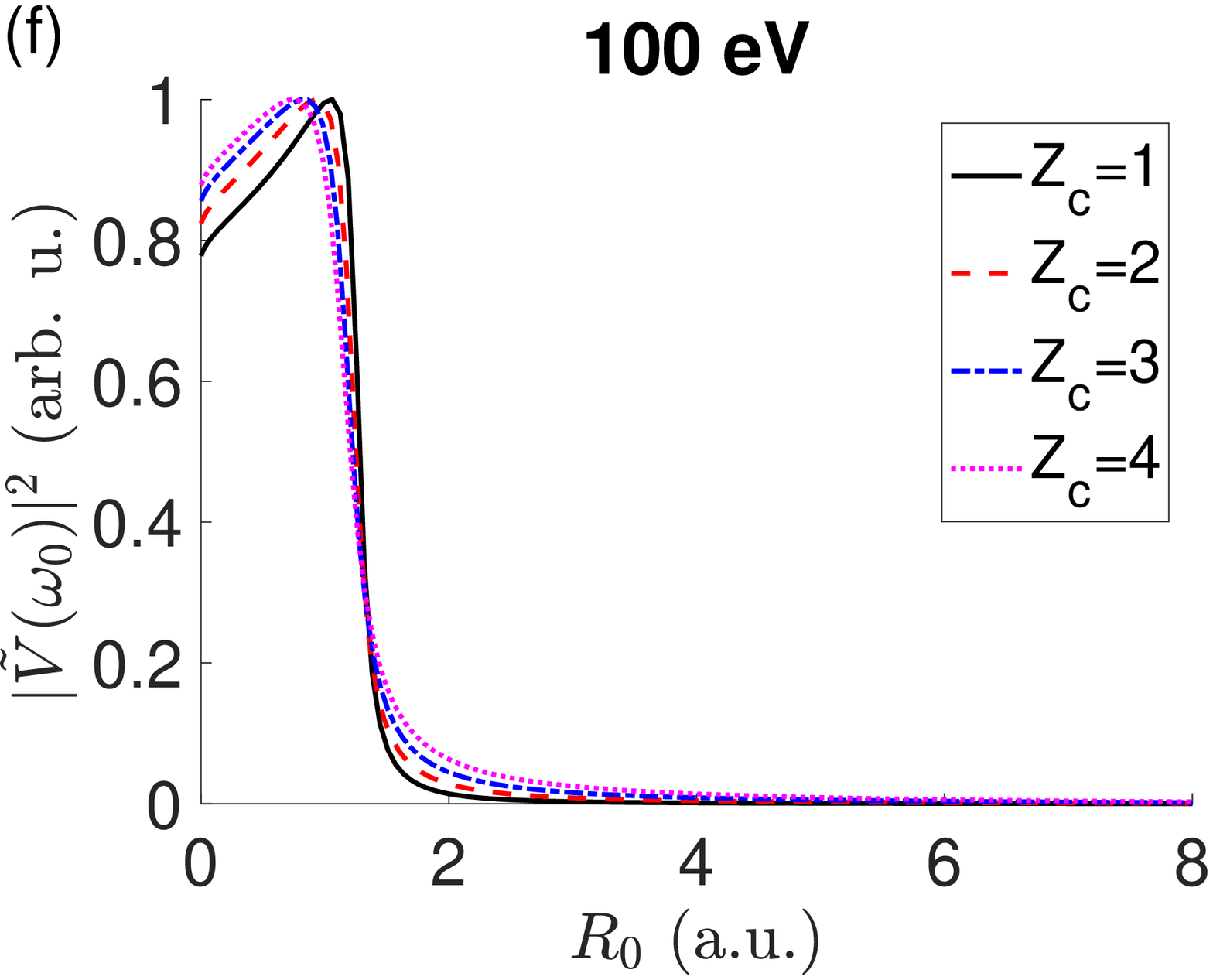}
 \caption{Dependency of $|\tilde{V}(\omega_0)|^2$ on $b$ (left column) and on $R_0$ (right column), for different electron energies 10, 50, and 100 eV, as labeled on figure. Each curve has been normalized to its own peak value. For each energy, four different ion-core charges have been used.}\label{f.Vom0}
\end{figure}

An electron collision trajectory is illustrated in Fig. \ref{f.bR0} (a). The electron has an asymptotic velocity $v_0$, assuming pointing to the $+x$ direction, and an impact parameter $b$. As the electron approaches the nucleus, it is accelerated by the ion-core potential but its total energy (kinetic energy plus potential energy) remains to be $v_0^2/2$. Denote $R_0$ be the electron-nucleus distance as the electron passes through the $x=0$ plane (the ``recollision plane" in the case of laser-driven recollision). Obviously $R_0 < b$. The detailed relationship between them depends on the value of $b$, the energy of the electron, and the form of the ion-core potential. Fig. \ref{f.bR0} (b-d) show the dependency of $R_0$ on $b$, for three different electron energies 10, 50, and 100 eV. For each energy, four different ion-core charges are used. For the ion-core potential, we have used the well-known Green-Sellin-Zachor (GSZ) effective potential \cite{GSZ}
\beq
V(r) = \frac{1}{r} \left[ -(Z-N) - \frac{N}{(e^{r/d}-1)\xi+1} \right],  \label{e.GSZ}
\eeq 
where $Z=90$ is the charge of the nucleus, $N$ (=89, 88, 87, ...) is the number of the remaining electrons in the ion core, $d=0.927$ a.u. and $\xi = 5.58$ a.u. are two parameters. The ion-core charge $Z_c = Z-N$.

As can be seen from Fig. \ref{f.bR0} (b-d), $R_0$ increases linearly with $b$ (with slope 1) for $b$ larger than a few atomic units. The difference between $R_0$ and $b$ becomes smaller as the energy of the electron increases, as can be expected. In the RINE process, we are concerned with $R_0$ smaller than 2 or 3 a.u., as explained below. The corresponding values of $b$ are mostly below about 5 a.u. 

If the distance $b$ (or $R_0$) is too large, then the interaction between the electron and the nucleus is too weak that the trajectory does not contribute to nuclear excitation. More precisely, the mutual potential changes with time too slowly that the trajectory is {\it adiabatic} with respect to the nuclear transition. Only when $b$ or $R_0$ is small enough does the corresponding trajectory contribute to nuclear excitation. This adiabaticity of a collision trajectory can be put in mathematical term by looking for the following Fourier component
\beq
\tilde{V} (\om_0) = \int_{-\infty}^{\infty} V(t) e^{-i\omega_0 t} dt,
\eeq
where $\omega_0 = 8.3$ eV is the energy gap between the two nuclear states, and $V(t) = V[r(t)]$ is the time-dependent (GSZ) potential between the electron and the nucleus following the trajectory $r(t)$ of the electron. From the semi-classical picture of Coulomb excitation, the nucleus is excited by the time-dependent potential supplied by the electron, and the above Fourier transform naturally arises if the nuclear excitation is calculated using time-dependent perturbation theory \cite{Alder-56}. In Fig. \ref{f.Vom0}, $|\tilde{V}(\omega_0)|^2$ is shown as a function of $b$ (left column) and as a function of $R_0$ (right column) for electron energies 10, 50, and 100 eV. For each energy, four different ion-core charges are shown, as labeled on figure.

One can see from Fig. \ref{f.Vom0} that for all the cases, $|\tilde{V}(\omega_0)|^2$ has a rather sharp cutoff beyond which it drops quickly to zero. This means that electron trajectories with $b$ or $R_0$ larger than the cutoff distances do not contribute to the nuclear excitation. Let us denote the cutoff distances to be $b_c$ and $R_c$. For 10 eV, $R_c$ is between 3.0 and 3.5 a.u., depending weakly on the ion-core charge $Z_c$. The corresponding $b_c$ is around 5 a.u. For 50 eV, $R_c$ is about 2.0 a.u. and the corresponding $b_c$ is about 2.5 a.u. For 100 eV, $R_c$ is about 1.5 a.u. and the corresponding $b_c$ is about 2.0 a.u. The area within the radius $R_c$ is the effective collision area for the purpose of nuclear excitation. The electron flux within the cross area $\pi R_c^2$ comes from infinity within a (slightly larger) cross area $\pi b_c^2$. In the recollision case, the electron flux does not come from infinity. Nevertheless, we can image that the electron flux {\it comes as if} from infinity with cross area $\pi b_c^2$.

\subsection{2.4 Tunneling ionization and recollision}

\begin{figure} [t!]
 \centering
 \includegraphics[width=4.2cm, trim=0 0 0 0]{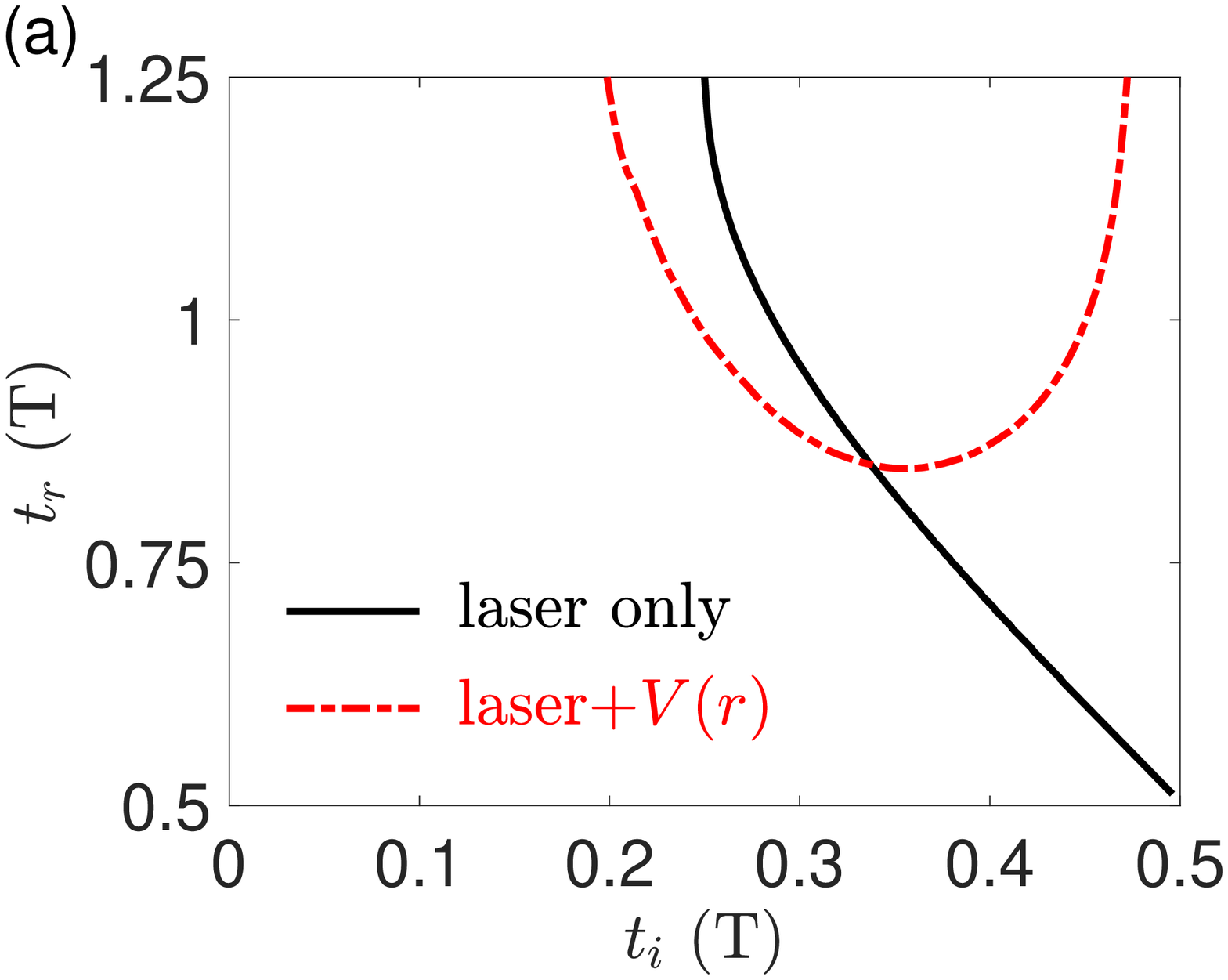} 
 \includegraphics[width=4.2cm, trim=0 0 0 0]{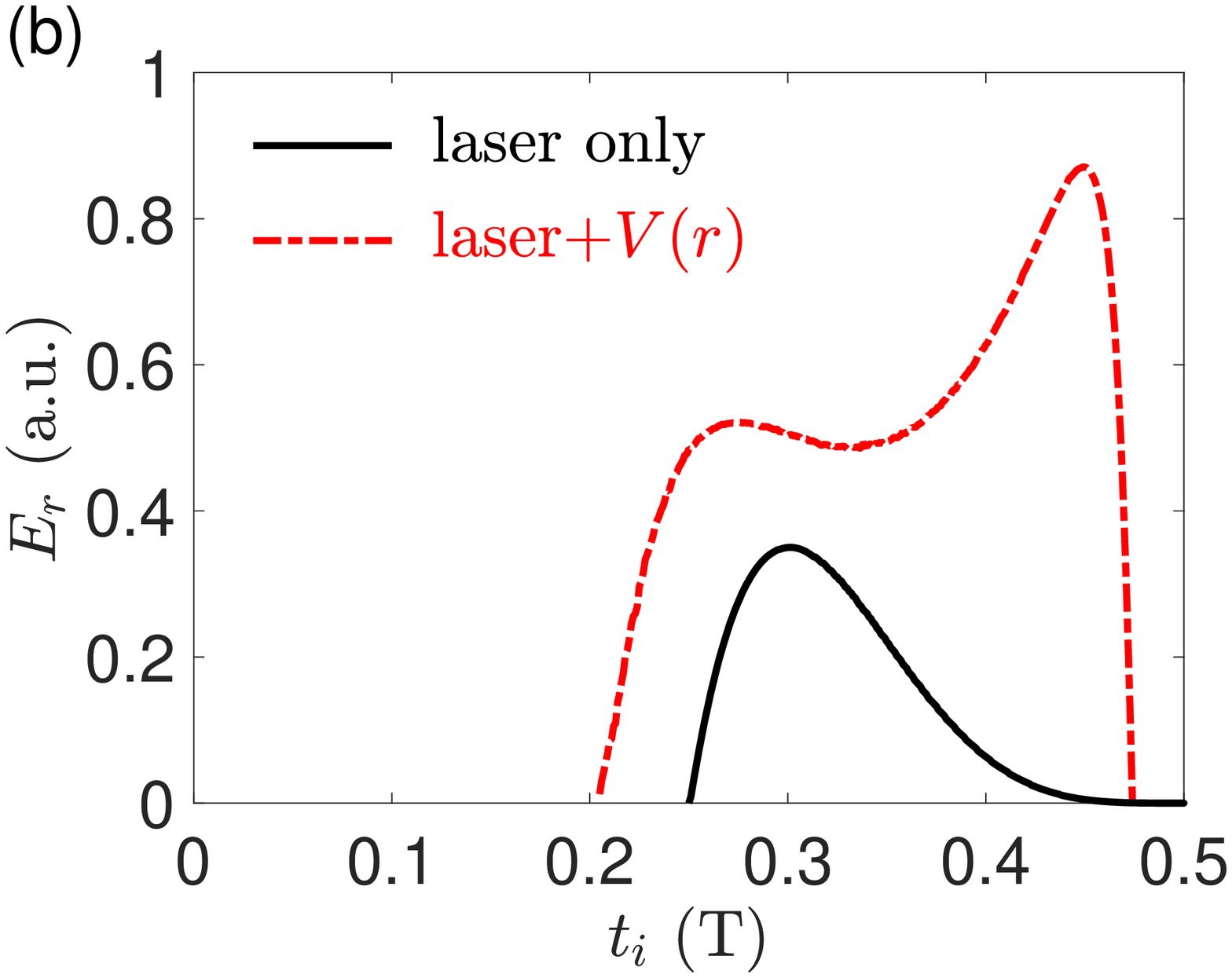} 
 \caption{(a) Relationship between the recollision time $t_r$ and the ionization time $t_i$. (b) Relationship between the recollision energy $E_r$ and the ionization time $t_i$. For both panels, the solid black curve is for the case where only the laser potential is considered, i.e. the Simpleman model, and the red dashed curve is for the case including the ion-core GSZ potential. The laser intensity used here is $5\times10^{13}$ W/cm$^2$ and the laser wavelength is 800 nm. }\label{f.recollision}
\end{figure}

A $^{229}$Th atom can be ionized, or even multiply ionized, in a strong laser field. The time-dependent ionization rate can be described by an Ammosov-Delone-Krainov (ADK) tunneling formula \cite{ADK}
\beq
w(t_i) =  \frac{f(l,m)}{\kappa^{2Z_c/\kappa-1}} \left( \frac{2\kappa^3}{|E(t_i)|} \right)^{2Z_c/\kappa-|m|-1} e^{-2\kappa^3/3|E(t_i)|}.    \label{e.ADK}
\eeq
Here $l$ and $m$ are the quantum numbers of the ionizing state, $\kappa \equiv \sqrt{2 I_{p}}$ with $I_{p}$ the ionization potential, $Z_c$ is the ion-core charge, and $E(t_i)$ is the laser electric field at the time $t_i$. The coefficient $f(l,m)$ is given by
\beq
f(l,m) = \frac{C^2_{l}}{2^{|m|} |m|!} \frac{(2l+1)(l+|m|)!}{2(l-|m|)!},
\eeq
where $C_l$ is a constant on the order of unity (in atomic units). Up to now no study has reported the values of $C_l$ particularly for $^{229}$Th, so we take $C_l = 1$ a.u. for the time being. Reported values of $C_l$ are mostly between 1 and 3 a.u. for rare gas atoms \cite{ADK}. The ionization probability is given by
\beq
P_\text{ion} (t_i) = 1 - \exp \left[ -\int_{t_0}^{t_i} w(t') dt' \right],  \label{e.Pion}
\eeq
where $t_0$ is the time when the laser pulse starts. The exponential term on the right hand side is the survival probability at the time $t_i$. 

The emitted electron has a chance to be driven back and recollide with its parent ion core when the oscillating laser electric field reveres its direction. Whether an electron recollides and with how much energy it recollides are determined by the emission time of the electron. In the simplest estimation one can ignore the ion-core potential and consider only the effect of the laser electric field. We assume that the electron is emitted at time $t_i$ at the position of the atom (taken as the origin) with zero initial velocity. Subsequent trajectory of the electron can be easily obtained. These assumptions constitute the so-called Simpleman model that has been very useful in qualitative understanding of strong-field phenomena \cite{Heuvell-88, Corkum-93}. 

Conclusions from the Simpleman model include: (i) The emitted electron can recollide if it is emitted after a laser field peak, e.g. if $0.25T \le t_i < 0.5T$ for a sinusoidal laser electric field $E(t) = E_0 \sin \omega t$. The electron cannot recollide if it is emitted before a field peak, e.g. if $0 < t_i < 0.25T$. Here $T = 2\pi/\omega$ is the laser period. The relationship between the ionization time $t_i$ and the corresponding recollision time $t_r$ is shown as the solid curve in Fig. \ref{f.recollision} (a). (ii) The maximum energy of the electron at the time of recollision is 3.17$U_p$ with $U_p = E_0^2/4\omega^2$ the ponderomotive potential. This maximum recollision energy is taken when the emission time $t_i = 0.3T$, i.e. 0.05 periods after the field peak. The dependency of the recollision energy $E_r$ on the emission time $t_i$ is shown as the solid curve in Fig. \ref{f.recollision} (b). 

The above conclusions are subject to modifications if the ion-core GSZ potential is included. For example: (i) Recollision may still happen if the ionization happens a little earlier than the peak of the laser electric field, as shown by the red dashed curve of Fig. \ref{f.recollision} (a). (ii) The recollision energy can be higher than the values from the Simpleman model, as shown by the red dashed curve of Fig. \ref{f.recollision} (b). 

If the ion-core GSZ potential is included, the initial position of the electron is set to be the tunneling-exit point, which can be solved by equating the total potential of the electron to the negative of the ionization energy on the polarization axis
\beq
V(r) + E(t_i) x = -I_p. \label{e.tunnelingexit}
\eeq
The initial momentum of the electron at the tunneling-exit point is usually assumed to be zero along the longitudinal direction \cite{Grochmalicki-91, Gajda-92, Chen-00, Yudin-01, Spanner-03, Quan-09, Smolarski-10, Ni-16, Pullen-17, Ivanov-17} (although some authors argue for slightly nonzero longitudinal momenta \cite{Pfeiffer-12, Camus-17, Tian-17, Wang-18, Xu-18}) and a Gaussian distribution along the transverse direction \cite{Ivanov-05}
\beq
P(v_\perp) \propto \exp\left(- \frac{v_\perp^2}{\eta^2} \right) \label{e.vperp}
\eeq
with $\eta^2 = |E(t_i)|/\sqrt{2I_{p}}$.

In the calculation the laser pulse duration is divided into small time steps with $dt_i = 0.005T.$ At each time step, $10^5$ trajectories are launched at the tunneling-exit point with random momenta along the transverse direction. Each trajectory is given a weight according to the above transverse-momentum distribution formula such that the total weight of the $10^5$ trajectories born at time $t_i$ is $w(t_i) dt_i [1-P_\text{ion} (t_i)]$. The value in the square bracket is the survival probability of the electron at the time. The convergence of the results has been checked by increasing the number of time steps and the number of trajectories at each time step.

\begin{figure} [t!]
 \centering
 \includegraphics[width=7cm, trim=0 0 0 0]{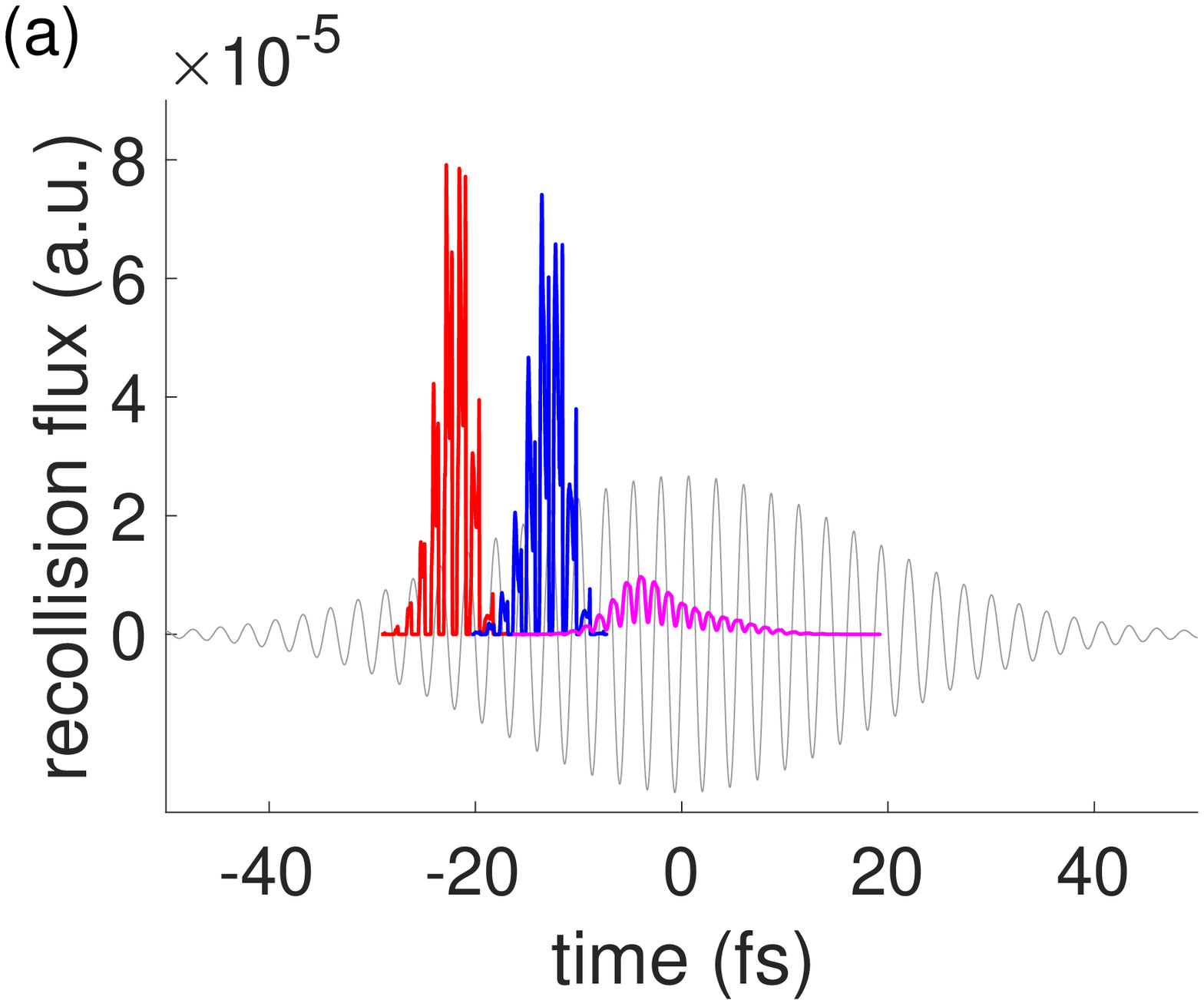} 
 \includegraphics[width=7cm, trim=0 0 0 0]{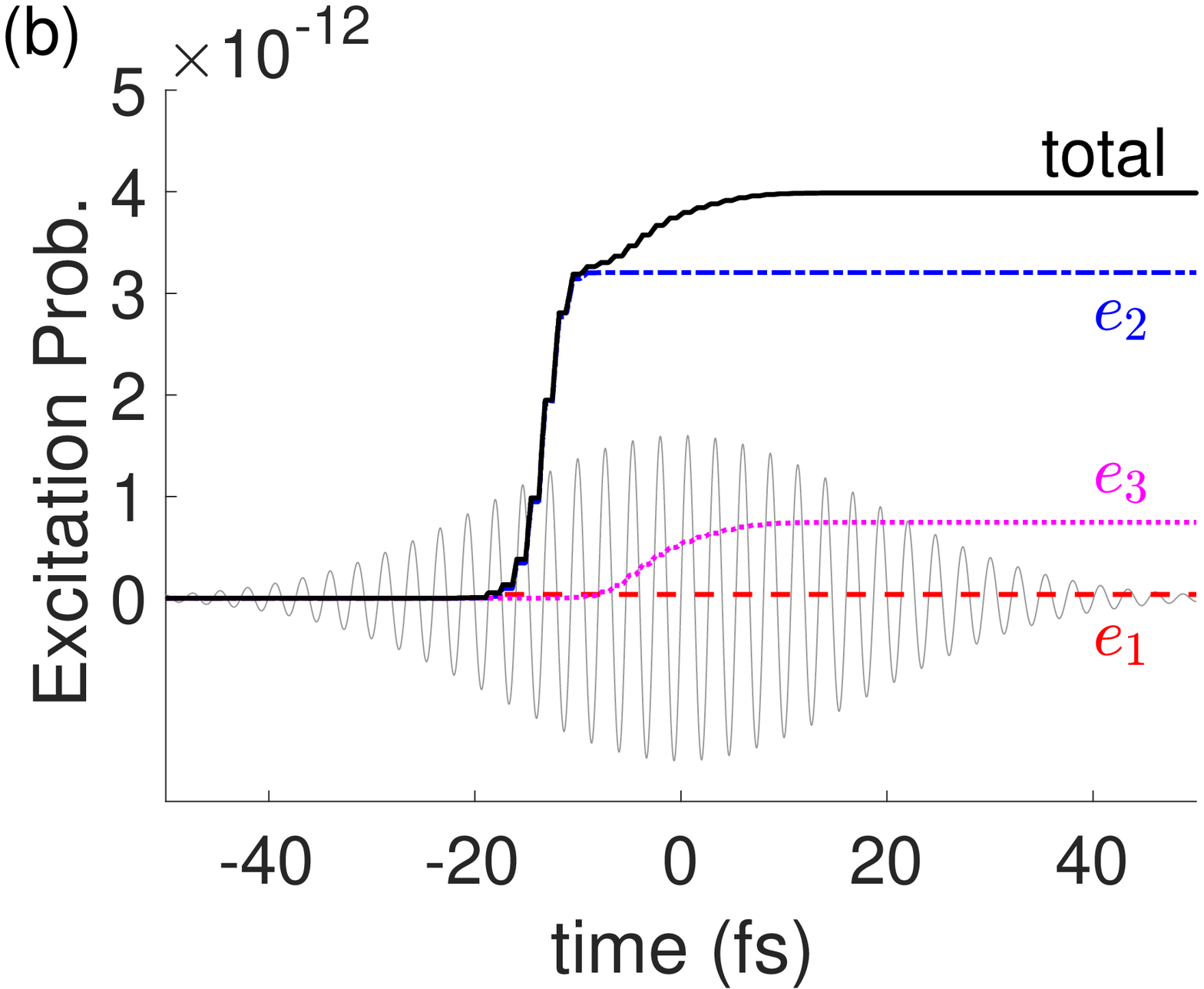}
 \caption{(a) Effective recollision flux density for the first-emitted electron (red, leftmost), the second-emitted electron (blue, middle), and the third-emitted electron (magenta, rightmost). The laser electric field is shown as the grey background. The pulse duration is 30 fs and the peak intensity is 10$^{14}$ W/cm$^2$. (b) The corresponding nuclear excitation probability. Both the total excitation probability and individual contribution from each electron are shown, as labeled.}\label{f.recflux}
\end{figure}

With the initial conditions specified, each trajectory is propagated according to the classical Hamiltonian equations of motion
\beqa
\frac{d r_i}{dt} &=& \frac{\partial H}{\partial p_i}; \\
\frac{d p_i}{dt} &=& -\frac{\partial H}{\partial r_i},
\eeqa
where $i=x,y,z$, and the Hamiltonian
\beq
H = \frac{1}{2}(p_x^2+p_y^2+p_z^2) + V(r) + xE(t). 
\eeq
The integrations are performed using the openly available LSODE (Livermore solver for ordinary differential equations) software package \cite{LSODE}. 
We follow each trajectory $r(t) = [x(t)^2+y(t)^2+z(t)^2]^{1/2}$ and determine whether it recollides (whether $x(t) = 0$ subsequently). If it does, the recolliding distance $R_0 = [y(t_r)^2 + z(t_r)^2]^{1/2}$ is recorded. As explained earlier, only collision trajectories with $R_0$ smaller than the critical distance $R_c$ contribute to nuclear excitation. The probability $P(t_i, R_c)$ in Eq. (\ref{e.jtr}) can be obtained by summing the weights of all the contributing trajectories born at $t_i$ and then dividing the total weight of all the trajectories born at the same time.

It is worth mentioning that similar tunneling-ionization-plus-classical-trajectory methods have been widely used in strong-field atomic physics to simulate strong laser-atom interactions \cite{Grochmalicki-91, Gajda-92, Chen-00, Yudin-01, Spanner-03, Quan-09, Smolarski-10, Ni-16, Pullen-17, Ivanov-17}.

\subsection{2.5 The effective recollision flux density}

The effective recollision flux density is calculated using Eq. (\ref{e.jtr}). Fig. \ref{f.recflux} (a) shows the recollision flux densities for a 30 fs (full width at half maximum, FWHM) Gaussian pulse with peak intensity 10$^{14}$ W/cm$^2$. Under this intensity, the outermost three electrons can be ionized (The first two electrons are completely ionized. The third electron has an ionization probability of 35\%. The ionization of the fourth electron is negligible). The recollision time intervals of the three electrons are separated, though, as can be seen from the figure. The first electron has the smallest ionization energy, so it is emitted and recollides early during the rising edge of the pulse. The second electron and the third electron follow. The effective recollision flux is calculated to be on the order of $10^{-5}$ a.u. 

More electrons can be emitted and contribute to nuclear excitation as the intensity increases, as will be shown later.

\begin{figure} [t!]
 \centering
 \includegraphics[width=7cm, trim=0 0 0 0]{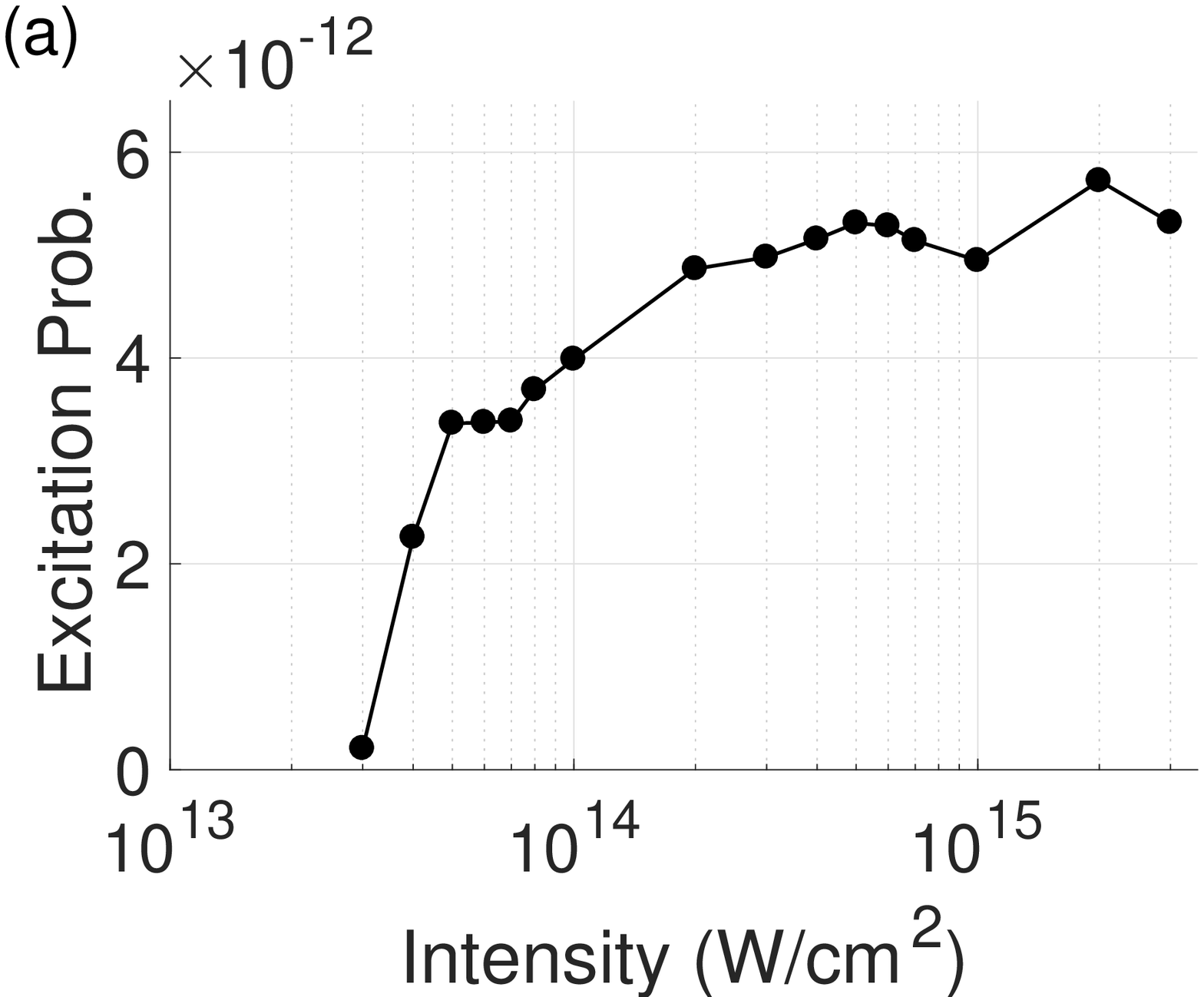}
 \includegraphics[width=7cm, trim=0 0 0 0]{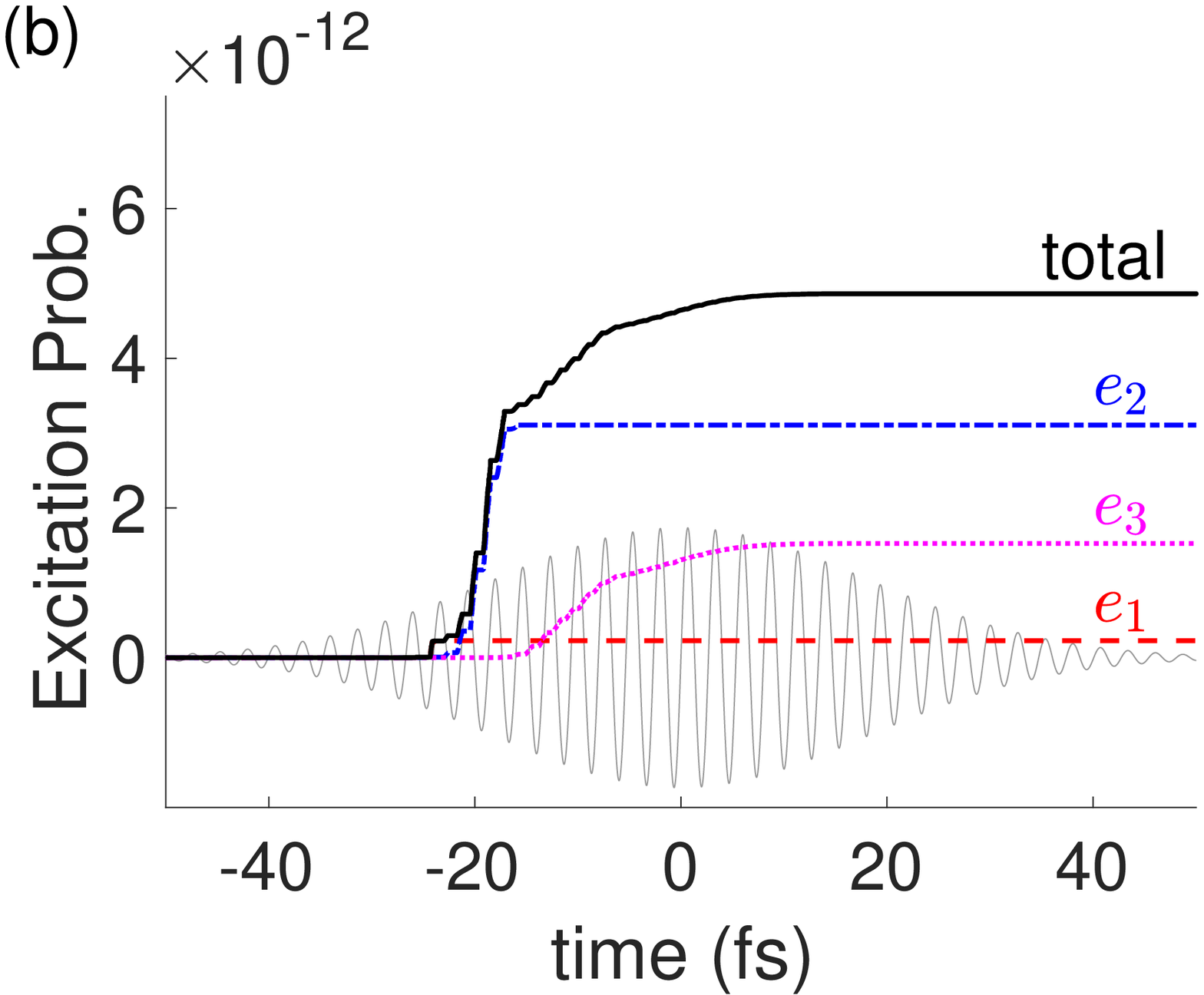}
 \includegraphics[width=7cm, trim=0 0 0 0]{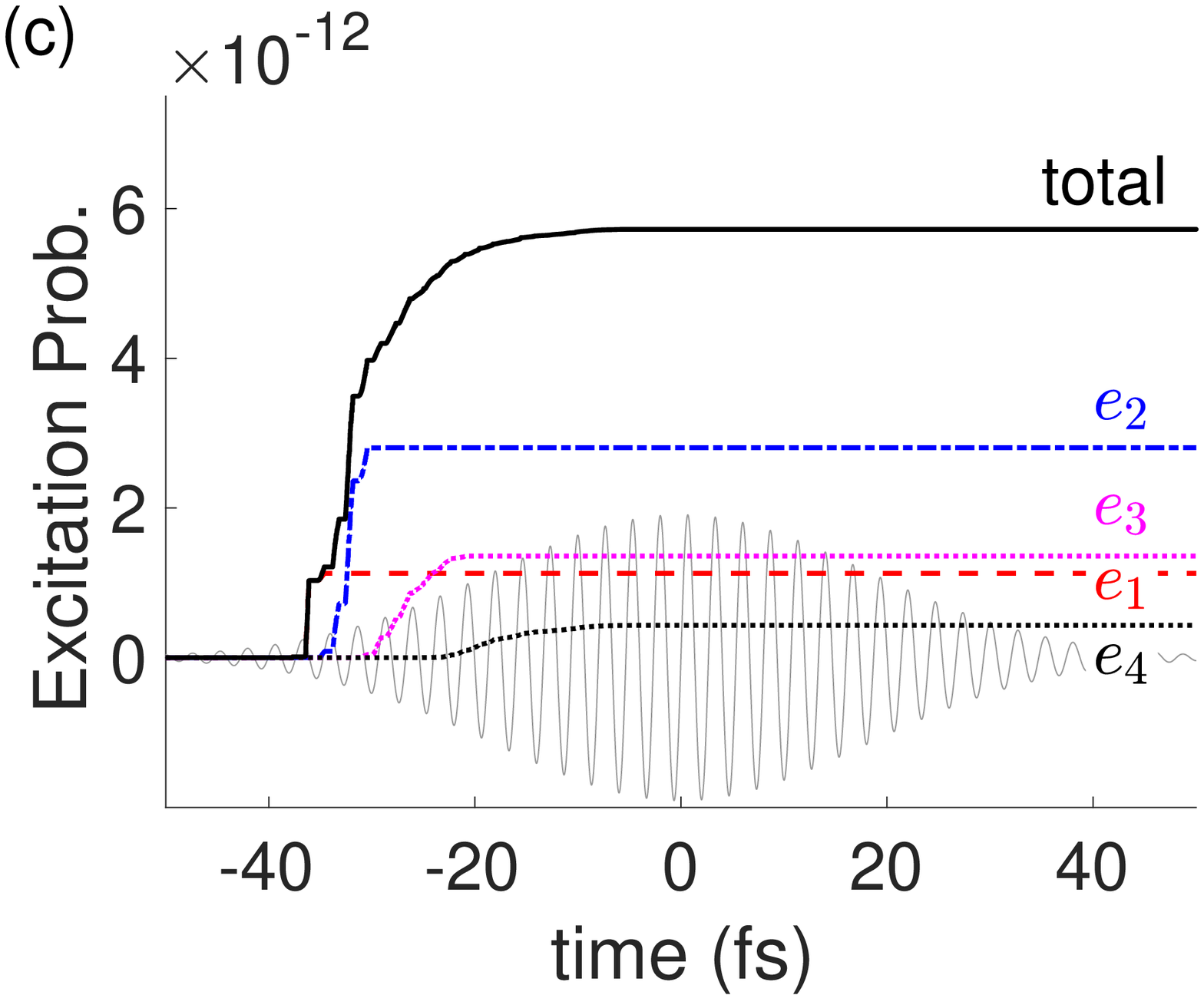}
 \caption{(a) Nuclear excitation probability as a function of laser peak intensity. The laser pulse has wavelength 800 nm and pulse duration 30 fs. (b) Accumulated excitation probability during the laser pulse for intensity $2\times10^{14}$ W/cm$^2$. Contributions from each electron are also shown, as labeled. (c) Similar as (b), but for intensity $2\times10^{15}$ W/cm$^2$.}\label{f.800inten}
\end{figure}

\subsection{2.6 The nuclear excitation probability}

The nuclear excitation probability is calculated using Eq. (\ref{e.Pexc}) for each contributing electron, and an example is shown in Fig. \ref{f.recflux} (b). Individual contribution from each electron to nuclear excitation is also shown. The excitation probability depends on both the effective flux density and the excitation cross section, which is determined by the electron energy at the recollision time. Note that set 1 of the reduced transition probabilities [Eqs. (\ref{e.BM1set1} - \ref{e.BE2set1})] have been used to calculate the cross section and the excitation probability.  

One can see from Fig. \ref{f.recflux} that although the first electron has higher flux densities, it does not contribute to nuclear excitation. This is because the recollision energy of the first electron is lower than the 8.3 eV excitation threshold. The first electron has a low ionization potential (6.3 eV), so it is emitted early during the rising edge of the laser pulse and experiences relatively weak laser fields, under which the recollision energies are not sufficient. This is of course a waste of electron fluxes. As will be shown later in Section 3.2, shorter laser pulses (with durations $\approx$ 10 fs) can increase the recollision energy of the first electron and make it useful in nuclear excitation.

\section{3. Numerical Results}

In this section extended numerical results are presented for the nuclear excitation probability. Dependencies on the laser intensity, the laser pulse duration, and the laser wavelength are calculated and analyzed.

\subsection{3.1 Dependency on laser intensity}

Fig. \ref{f.800inten} (a) shows the dependency of the (end-of-pulse) nuclear excitation probability on the laser peak intensity. The pulse duration is fixed at 30 fs and the laser wavelength is 800 nm. One can see that there is a threshold intensity around $3\times 10^{13}$ W/cm$^2$ below which nuclear excitation cannot happen via recollision. This is because the recollision energy of the electron cannot reach the 8.3 eV excitation threshold below this intensity. Above the threshold intensity the excitation probability quickly increases. The excitation probability is on the order of $10^{-12}$ and increases as a general trend with intensity. 

Fig. \ref{f.800inten} (b) and (c) show the accumulation of the excitation probability during the laser pulse for two different laser intensities, namely, $2\times10^{14}$ and $2\times10^{15}$ W/cm$^2$. For the former intensity, three electrons are emitted and contribute to the nuclear excitation. Comparing to the case of $1\times10^{14}$ W/cm$^2$ as shown in Fig. \ref{f.recflux} (b), the first-emitted electron now has a small but recognizable contribution to nuclear excitation. For the latter intensity, four electrons are emitted and contribute to the nuclear excitation. If the laser intensity increases further, more electrons can be emitted and contribute to nuclear excitation.

\subsection{3.2 Dependency on laser pulse duration}

\begin{figure} [t!]
 \centering
 \includegraphics[width=7cm, trim=0 0 0 0]{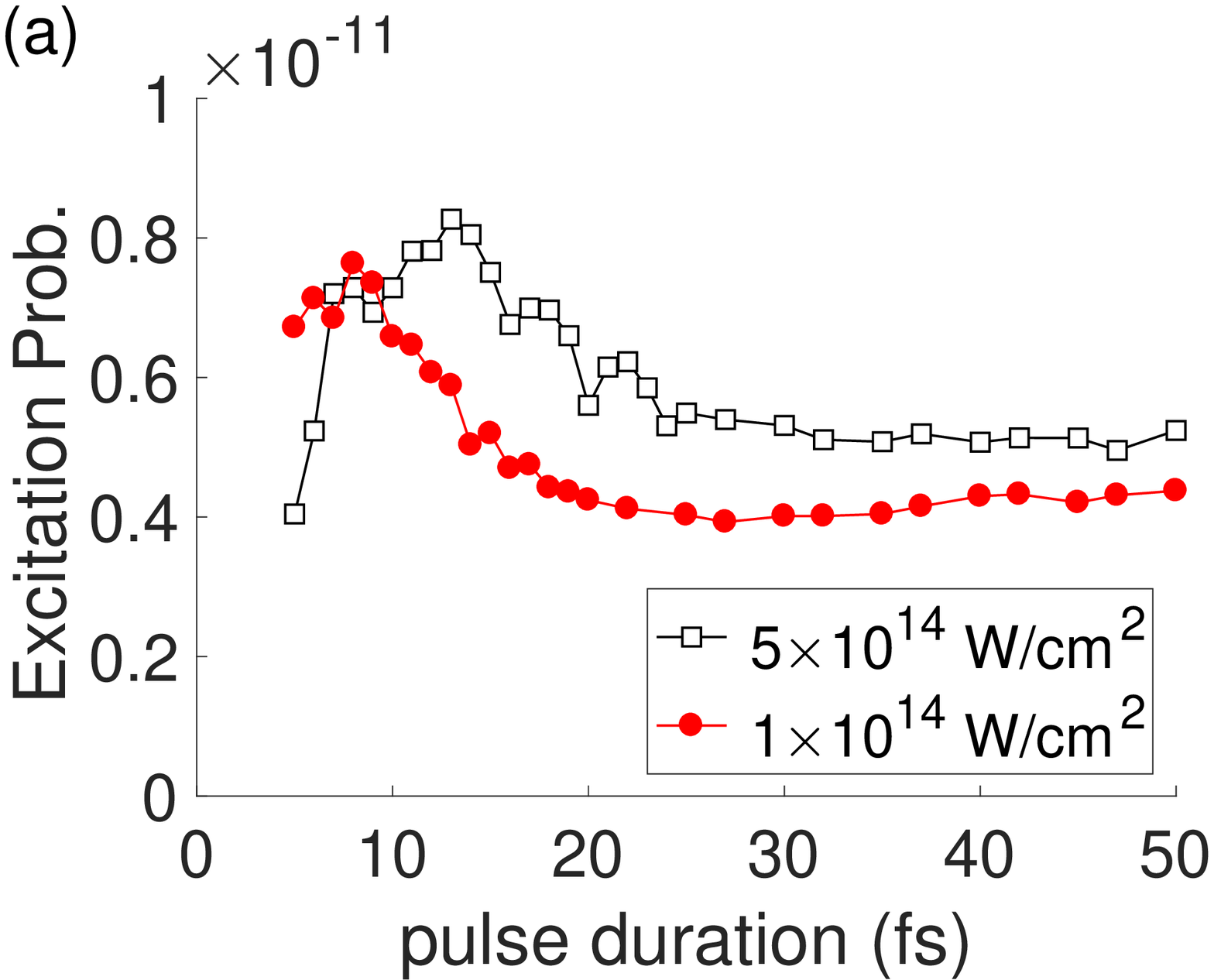} 
 \includegraphics[width=7cm, trim=0 0 0 0]{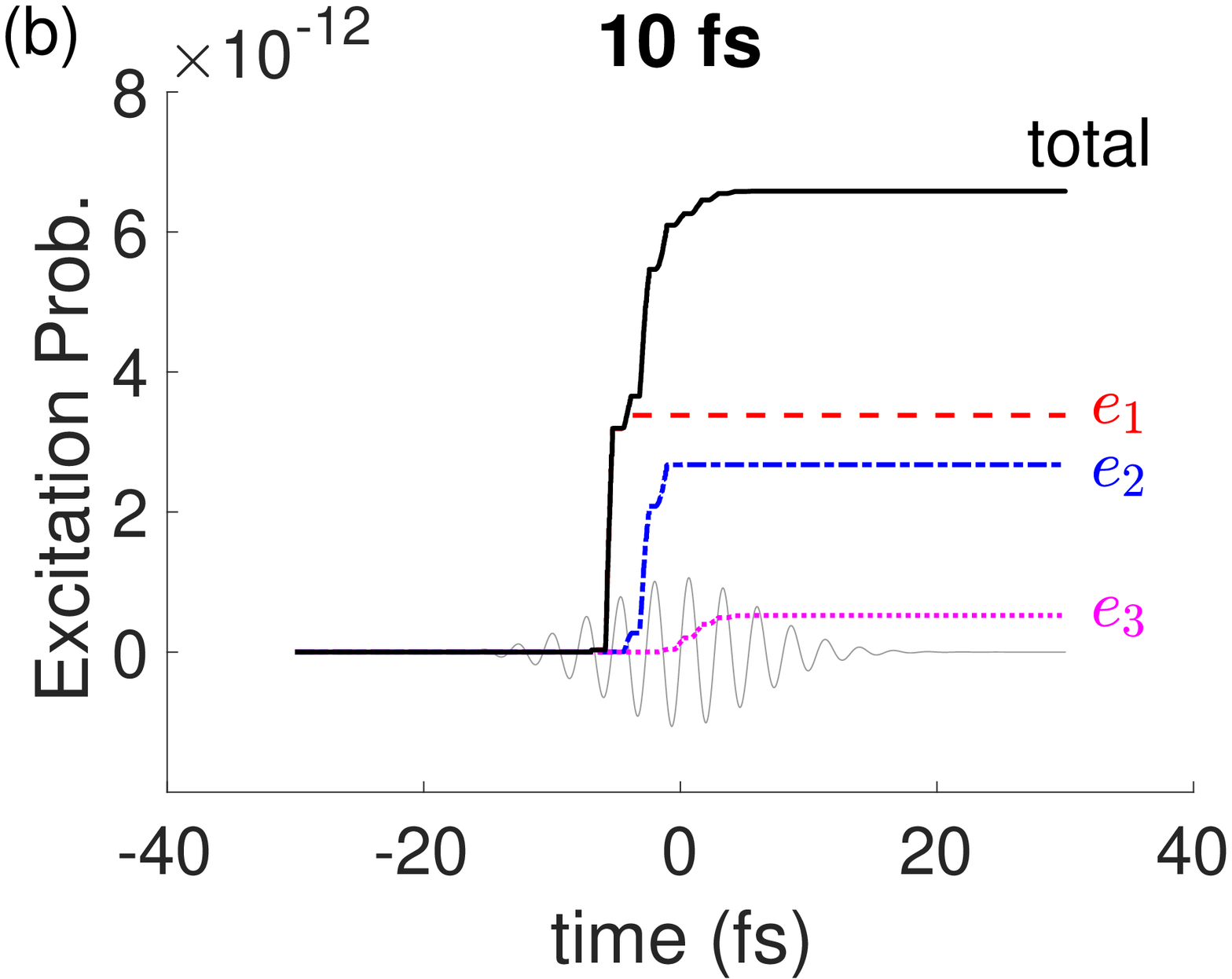}
 \includegraphics[width=7cm, trim=0 0 0 0]{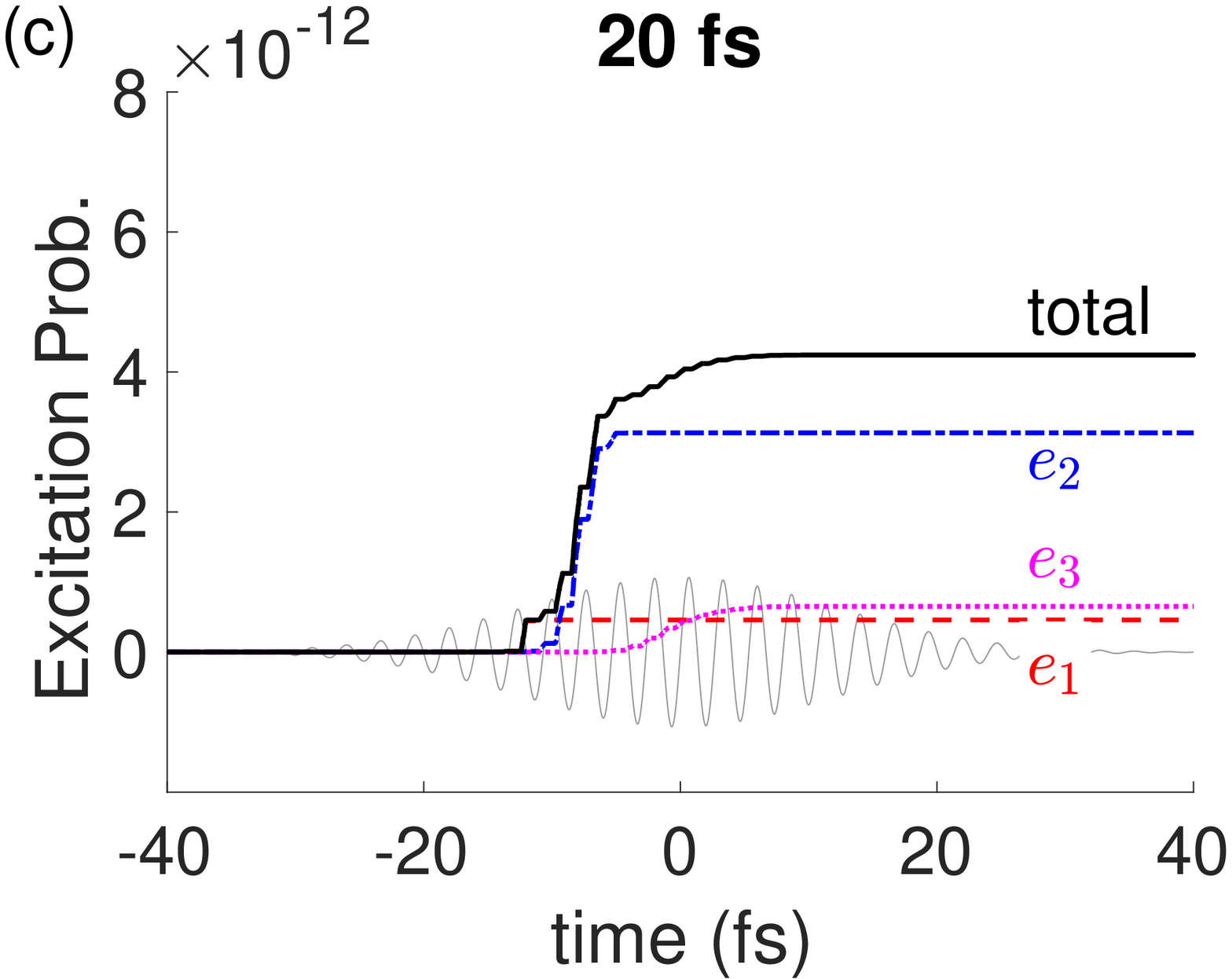}
 \caption{(a) Nuclear excitation probability as a function of laser pulse duration, for two different peak intensities as labeled on figure. The laser pulses have wavelength 800 nm. (b) Accumulated excitation probability during a laser pulse for intensity $1\times10^{14}$ W/cm$^2$ and pulse duration 10 fs. (c) Similar as (b), but for pulse duration 20 fs.}\label{f.800pulsedur}
\end{figure}

Fig. \ref{f.800pulsedur} (a) shows the dependency of nuclear excitation probability on the laser pulse duration, for two difference intensities as labeled. The laser wavelength is fixed at 800 nm. For both intensities, one can see that the excitation probability is rather flat for pulse durations longer than about 20 fs. Shorter pulses lead to higher excitation probabilities. The excitation probability can be about twice as high compared to the long-pulse values.

The increased excitation probability for shorter pulses mainly comes from the activation of the first-emitted electron. Fig. \ref{f.800pulsedur} (b) and (c) show the nuclear excitation probability during two pulses of different durations, namely, 10 fs and 20 fs. Both pulses have intensity $1\times10^{14}$ W/cm$^2$ and wavelength 800 nm. One can find that the contributions from the second and the third electrons are almost the same for the two pulse durations. The difference comes from the first electron. For the 10 fs case, the first electron contributes the most to nuclear excitation among the three electrons. Whereas for the 20 fs case, the first electron contributes the least among the three electrons.

The above results can be understood as follows. Recollision is largely a process that happens within a laser cycle. Ionization happens around a field peak and recollision happens roughly $3/4$ cycles later around field zero. As long as the pulse is long enough, such that the pulse envelope does not change substantially within a laser cycle, the ionization-recollision process is similar from cycle to cycle. This explains the nearly flat behavior above 20 fs. However, if the pulse is very short so that the pulse envelope changes appreciably within a laser cycle, then the ionization-recollision behavior can be greatly modified. An emitted electron can feel quite a different electric field compared to a pure sinusoidal field, and its recollision time and recollision energy can be significantly modified. Interestingly, as the pulse becomes shorter, the recollision energy of the first-emitted electron increases to values above the 8.3 eV threshold and the first electron is ``activated" to excite the nucleus. This is good news since the flux density of the first electron is usually larger than other electrons.

Pulses durations around 10 fs seem to be the most efficient for the purpose of nuclear excitation. As the pulse duration decreases further, the ionization is suppressed, which reduces the recollision flux hence the nuclear excitation probability.

\subsection{3.3 Dependency on laser wavelength}

\begin{figure} [t!]
 \centering
 \includegraphics[width=7cm, trim=0 0 0 0]{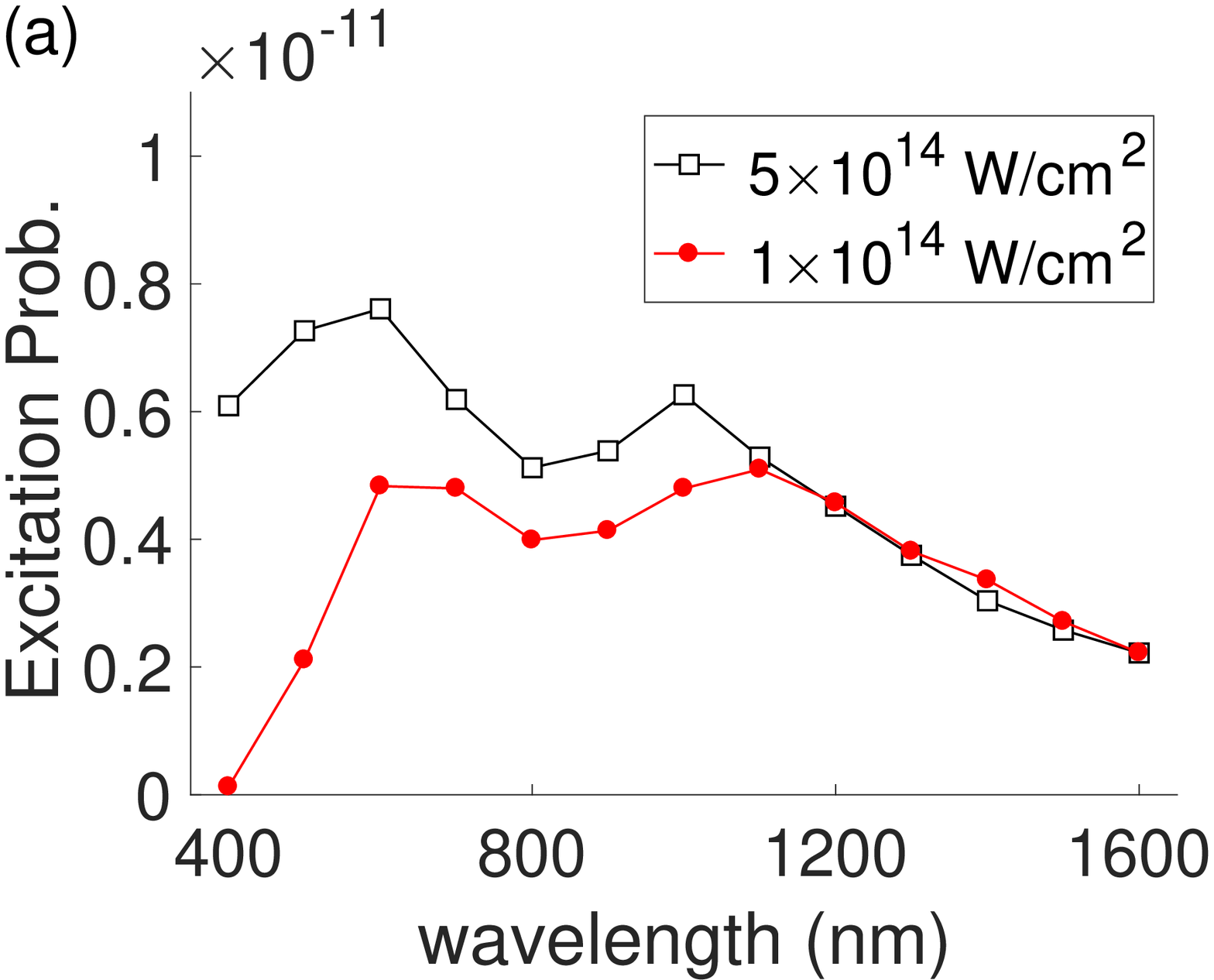} 
 \includegraphics[width=7cm, trim=0 0 0 0]{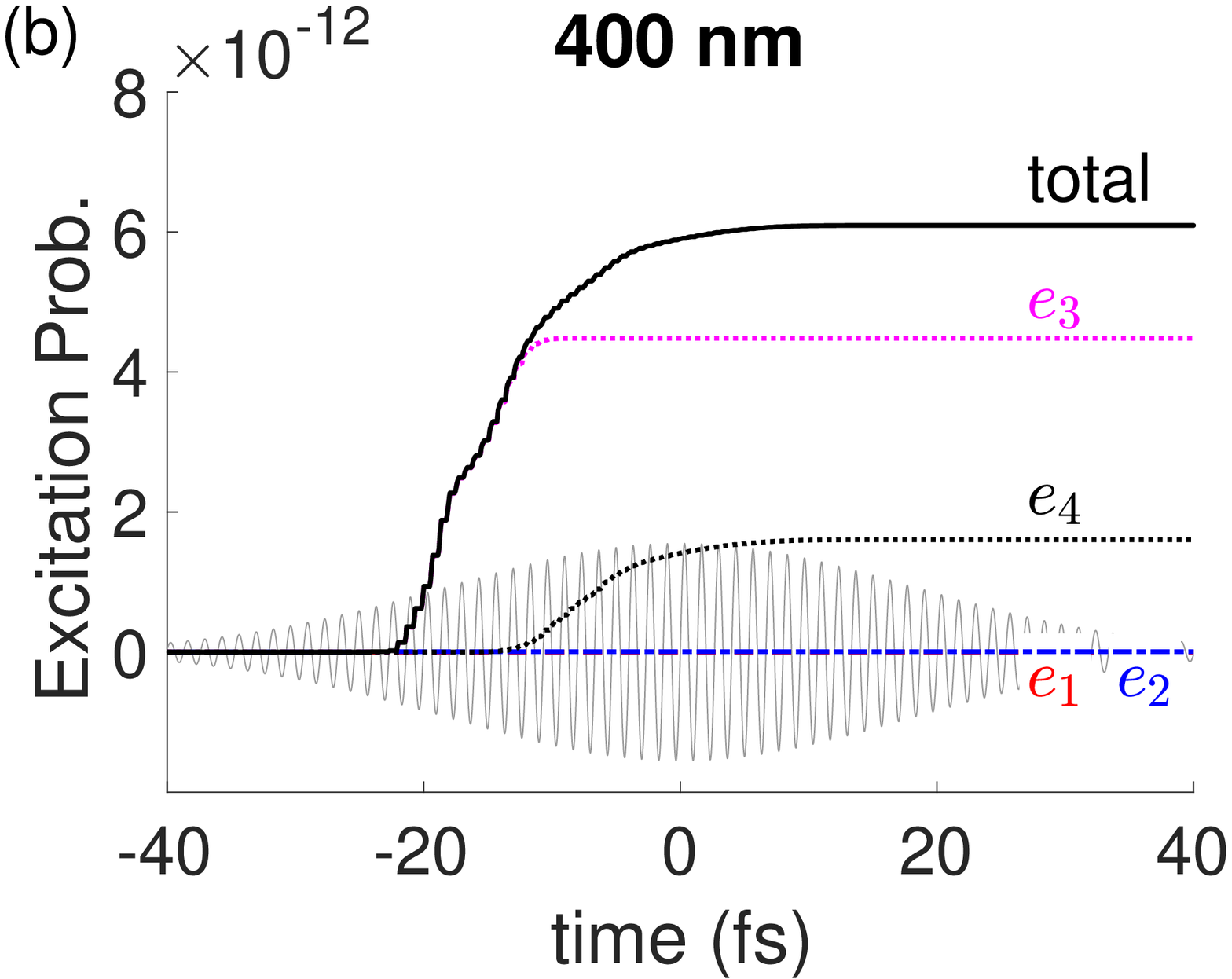}
 \includegraphics[width=7cm, trim=0 0 0 0]{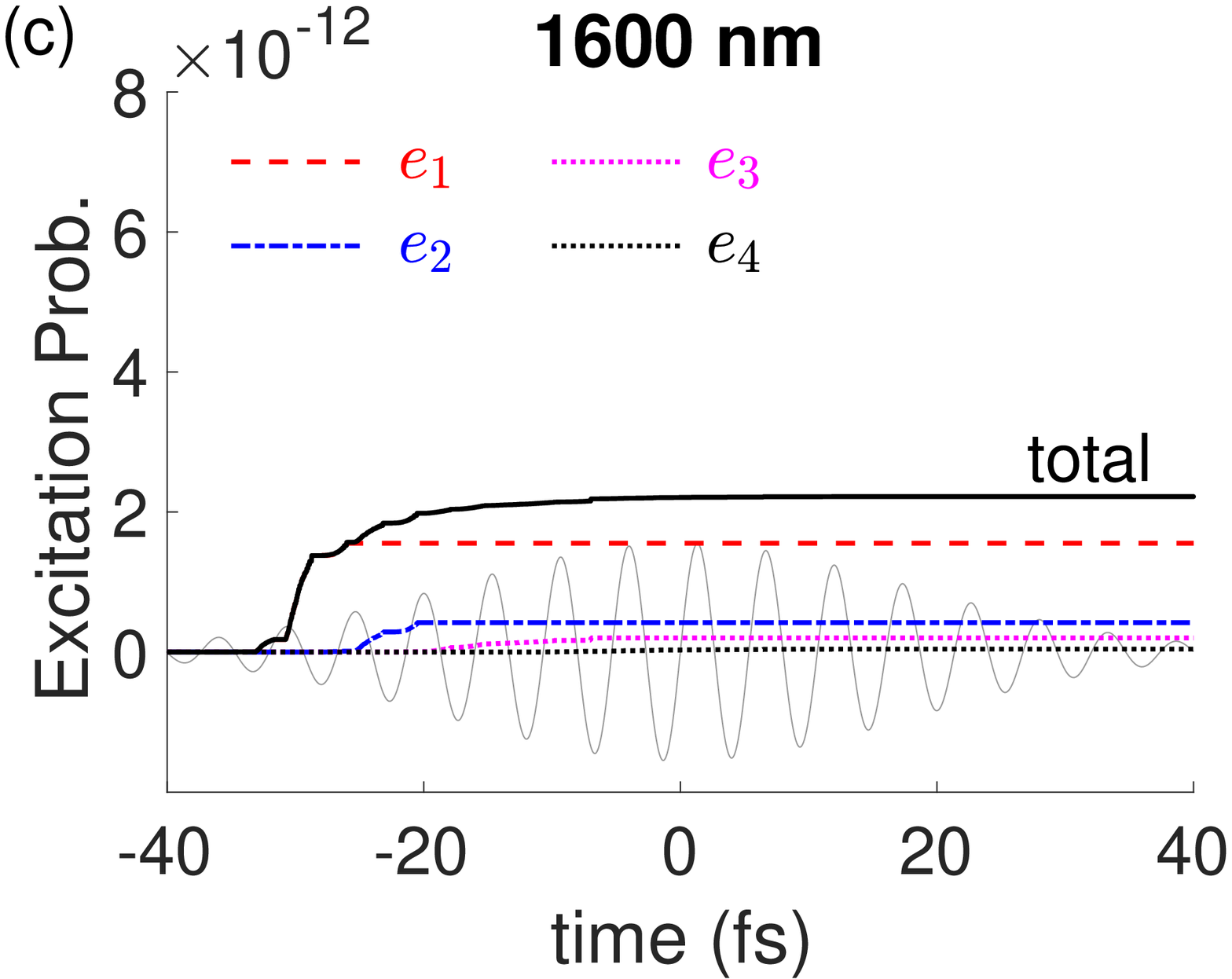}
 \caption{(a) Nuclear excitation probability as a function of laser wavelength. The pulse duration is fixed at 30 fs. Two peak intensities are shown, as labeled on figure. (b) Accumulated excitation probability during a laser pulse of wavelength 400 nm and intensity $5\times10^{14}$ W/cm$^2$. (c) Similar as (b), but for wavelength 1600 nm.}\label{f.wavelength}
\end{figure}

Fig. \ref{f.wavelength} (a) shows the dependency of the nuclear excitation probability on the laser wavelength, within a range from 400 to 1600 nm. The laser pulse duration is fixed at 30 fs. Results for two peak intensities are shown, as labeled on figure. 

The laser wavelength has twofold effects on the recollision process. First, longer wavelengths lead to higher recollision energies. Second, longer wavelengths lead to more severe wavepacket spreading along the transverse direction, hence lower recollision fluxes. These two aspects manifest in Fig. \ref{f.wavelength} (a). For example, the excitation probability decreases for wavelengths longer than about 1000 nm. This is a manifestation of transverse wavepacket spreading. For another example, the excitation probability also drops as the wavelength becomes shorter than 600 nm. This is a manifestation of reduced recollision energies as the wavelength decreases. 

Fig. \ref{f.wavelength} (b) and (c) show two examples of accumulated nuclear excitation probabilities during the laser pulse. For the 400 nm case shown in (b), the first electron and the second electron do not contribute to nuclear excitation. The recollision energies are below the 8.3 eV threshold albeit with a relatively high laser intensity of $5\times10^{14}$ W/cm$^2$. Fortunately this intensity is strong enough to pull out four electrons, and the third and the fourth electrons have enough recollision energies to excite the nucleus. If the intensity decreases to a value with which the third electron and the fourth electron barely ionize, then the nuclear excitation probability will be greatly suppressed. This is the case for intensity $1\times10^{14}$ W/cm$^2$. For the 1600 nm case shown in (c), the first electron has enough recollision energy to excite the nucleus. However, the excitation probabilities decrease substantially for the second, the third, and the fourth electrons, due to severe wavepacket spreadings along the the transverse direction. The laser wavelength therefore is quite an efficient knob to control recollision and the nuclear excitation process.

\section{4. Discussions}

\subsection{4.1 Effects of the ion-core potential (Coulomb focusing)}

\begin{figure} [t!]
 \centering
 \includegraphics[width=7cm, trim=0 0 0 0]{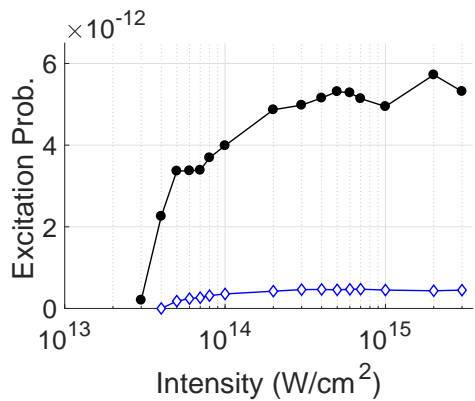} 
 \caption{Nuclear excitation probability with (black filled circles) and without (blue unfilled diamands) the ion-core GSZ potential. The black curve with filled circles is the same as Fig. \ref{f.800inten} (a). The laser wavelength is 800 nm and the pulse duration is 30 fs.}\label{f.Coulombfocusing}
\end{figure}

If the ion-core GSZ potential is removed from the calculation (the electron trajectories are propagated under the driven of only the laser electric field), then the nuclear excitation probability is found to drop by over an order of magnitude, as shown in Fig. \ref{f.Coulombfocusing}. The ion-core potential plays an important role in focusing the electron trajectories along the transverse direction and enhancing the effective recollision flux density. This effect is called Coulomb focusing which is well known in strong-field atomic physics \cite{Brabec-96, Comtois-05, Shafir-13, Danek-18}.

\subsection{4.2 Loss of flux as the recolliding electron travels through the ion-core electron cloud}

The recolliding electron penetrates through the electron cloud of the remaining ion core before getting close enough to excite the nucleus. During this process the recolliding electron may interact with the ion-core electrons and loses part of its flux. It is therefore important to estimate this loss.

\begin{figure} [t!]
 \centering
 \includegraphics[width=7cm, trim=0 0 0 0]{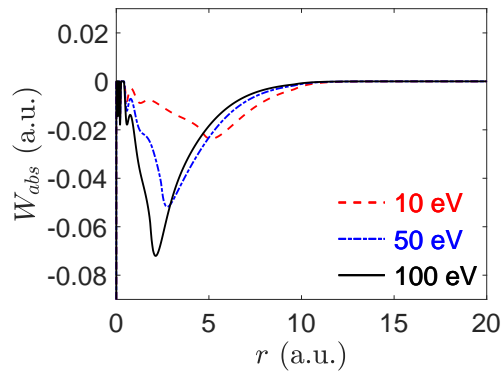} 
 \caption{Absorption potential felt by the recolliding electron as it travels through the electron cloud of $^{229}$Th$^{+}$. The potential is calculated using the code {\tt ELSEPA} with three different electron incoming energies as labeled.}\label{f.Wabs}
\end{figure}

Here we use the code {\tt ELSEPA} \cite{ELSEPA} to calculate the imaginary absorption potential $iW_\text{abs}(r)$ of the $^{229}$Th$^{+}$ ion, felt by an electron coming close to it. The absorption potential is shown in Fig. \ref{f.Wabs}, for three different electron incoming (asymptotic) energies. The probability that the electron is lost from the flux (absorbed) can be estimated to be
\beq
P_\text{abs} = 1 - \exp\left[\int_0^{\infty} 2 W_\text{abs}(r)\frac{dr}{v(r)}\right],
\eeq
where $v(r)=\sqrt{2(E_i+1/r)}$ is the electron velocity at distance $r$. In writing the above formula we have, for simplicity, assumed a straight trajectory passing through the electron cloud on the $x$ axis. Trajectories off the axis have shorter traces inside the electron cloud so they are expected to be less absorbed than the on-axis case. $P_\text{abs}$ is calculated to be about 19\% for 10 eV, 16\% for 50 eV, and 14\% for 100 eV. These results change very little for the $^{229}$Th$^{2+}$, $^{229}$Th$^{3+}$, and the $^{229}$Th$^{4+}$ ion cores. Based on these results, we conclude that the recolliding electrons indeed lose some fluxes, but the majority of the fluxes can penetrate the ion-core electron cloud and contribute to nuclear excitation.

\subsection{4.3 Advantages of the RINE approach}

Compared with other excitation approaches, the RINE approach has the following advantages, some of which are unique:

(i) It is efficient. The probability for a single $^{229}$Th nucleus to be excited during a 10 fs pulse is calculated to be on the order of $10^{-12}$. In comparison, for the 29 keV indirect light excitation approach, which is the only experimentally realized excitation approach so far, the excitation probability for a single $^{229}$Th nucleus is estimated to be $10^{-11}$ per second \cite{Masuda-19}. That is, the same excitation probability can be achieved with RINE during a 10 fs laser pulse as the indirect light excitation approach does during 0.1 seconds. 

(ii) A precise knowledge of the isomeric energy is not needed. The current knowledge of the isomeric state energy of 8.3 eV ($8.28\pm 0.17$ eV as given in Ref. \cite{Seiferle-19}) is still uncertain by a fraction of an electronvolt, which leads to troubles for excitation approaches requiring a precise knowledge of it. This is not a problem for the RINE approach because the recolliding electrons have a range of energies instead of a single energy. The isomeric energy is certainly covered by the energy range of the recolliding electrons.

(iii) Large light facilities like synchrotron radiations are not needed. The RINE approach only needs table-top femtosecond laser systems which are much more accessible. 

(iv) The nuclear excitation is well timed. All the excitations happen within (in fact, a fraction of) the femtosecond laser pulse, instead of distributing over all the time. This might be a crucial property for future coherent operations of the isomeric state.

(v) The excited nuclei are accompanied with well controlled ionic states, and they will not decay via internal conversion. The resultant $^{229}$Th ions have almost no recoil energies, unlike those from the decay of $^{233}$U.

\subsection{4.4 Further remarks}

(i) We have used a tunneling-ionization-pulse-classical-trajectory method to describe the ionization and the recollision processes. This method is physically very intuitive and computationally moderate. A more fundamental (quantum mechanical) method would be to solve the time-dependent Schr\"odinger equation for the ionization and subsequent evolution of an electron. This can be done under a single-active-electron approximation, but the computational load will be much higher and the physics less intuitive.

(ii) We have considered only first-order recollisions and neglected higher-order ones. This should be well justified because higher-order recollisions associate with much more severe wavepacket spreadings along the transverse direction, hence much lower recollision fluxes.

(iii) We have calculated the ionization rates using the ADK tunneling formula, which is valid in the tunneling regime. A possible extension would be to use the Perelomov-Popov-Terent'ev (PPT) formula \cite{PPT}, which is applicable also to the multiphoton regime \cite{Keldysh-65}.

\section{5. Conclusion}

In this article I have elaborated our previously proposed RINE approach \cite{Wang-21} for the excitation of $^{229}$Th nucleus. The method itself has been explained in further details, with the involved elements carefully examined, including the electronic excitation cross section, the adiabaticity of collision trajectories and the critical collision radius, the ionization and the recollision processes, the effective recollision flux density, etc. I believe that the RINE approach is now more solidly founded. 

Numerical results show that the RINE approach leads to nuclear isomeric excitation probabilities on the order of $10^{-12}$ per nucleus per (femtosecond) laser pulse. Dependency of the nuclear excitation probability on the laser intensity, the laser pulse duration, and the laser wavelength has been calculated and analyzed. The nuclear excitation process can be efficiently controlled by varying these laser parameters. Additional discussions have also been made on the effect of the ion-core Coulomb focusing and on the loss of recollision flux when the recolliding electron flies through the ion-core electron cloud.

$^{229}$Th is a fascinating system with important potential applications. Apart from the applications, it also provides an interesting platform on which nuclear physics, atomic physics, and laser physics directly interplay. An example of such an interplay is the RINE method, which combines $^{229}$Th nuclear physics with strong-field atomic physics. It would certainly be interesting to see new manifestations of this three-partite interplay.

I acknowledge discussions with Mrs. Hanxu Zhang and Wu Wang, and funding support from Science Challenge Project of China No. TZ2018005, NSFC No. 12088101, and NSAF No. U1930403.

\end{document}